\documentclass[12pt]{article}

\makeatletter
\renewcommand{\theequation}{\arabic{section}.\arabic{equation}}
\makeatother

\newcounter{prop}[section]
\newcounter{theor}[section]
\newcounter{defin}[section]
\newcounter{theor_append}

\newenvironment{prop}
{
\refstepcounter{prop}%
{\ \newline
\bf Proposition
\arabic{section}.\arabic{prop}.}\em}
{}
\newenvironment{theor}
{
\refstepcounter{theor}%
{
\ \newline
\bf Theorem
\arabic{section}.\arabic{theor}.}\em}
{}
\newenvironment{defin}
{
\refstepcounter{defin}%
{\ \newline
\bf Definition
\arabic{section}.\arabic{defin}.}\em}
{}
\newenvironment{theor_append}
{
\refstepcounter{theor_append}%
{\ \newline
\bf Theorem
A.\arabic{theor_append}.}\em}
{}

\newcommand{\prf}{{\ \newline \bf Proof.}\ }
\newcommand{\eprf}{$ \triangle $}

\begin{document}

\title{Exactly solvable models of quantum mechanics including
fluctuations in the framework of representation of the wave function
by random process}
\author{A.S.Gevorkyan \and A.A.Udalov}
\date{Institute for High-Performance Computing and Data Bases,
P/O Box 71, 194291, St.Petersburg, Russia\\
ashot@fn.csa.ru\quad  udalov@fn.csa.ru}

\maketitle

\begin{abstract}
The problem of quantum harmonic oscillator with "regular+random"
square frequency, subjected to "regular+random external force, is
considered in framework of representation of the wave function by complex-valued
random process. Average transition probabilities are calculated.
Stochastic density matrix method is developed, which is used for
investigation of thermodinamical characteristics of the system, such as
entropy and average energy.
\end{abstract}


\section*{Introduction}
\label{sec-1}

Recently there have been published a great amount of papers
\cite{proc} concerned with the "quantum chaos", i.e. with the
quantum analogues of classical systems possessing the dynamic chaos
features. The investigations are conducted along the different
directions, such as analysis of the energy levels distribution;
definition and calculation of quantities which are responsible in
the quantum systems for the presence of chaos (corresponding to the
classical Lyapunov exponents and KS-entropy); study of localization
and delocalization of wave functions around classical orbits; etc.
It is worth saying that though in most of the cases mentioned above
one is faced with the necessity to describe a quantum system
statistically, so far there was not paid much attention to a
stochastic behavior of the wave function itself.

Many problems of great importance in the field of the
nonrelativistic quantum mechanics, such as description of Lamb
shift, spontaneous transitions in atoms, etc., remain unsolved due
to the fact that the concept of physical vacuum has not been
considered within the framework of the standard quantum mechanics.
It is obvious that a quantum object immersed into the physical
vacuum is an open system. There exist various approaches \cite{pot}
to the description of such systems, mainly in application to the
problem of continuous measurements. One of them is based on the
consideration of the wave function as a random process, for which a
stochastic differential equation (SDE) is derived. But the equation
is obtained by the method which is extremely
difficult for application even in case of comparatively simple type of interaction
between the system and the environment, so that some new ideas are
needed \cite{gc}-\cite{sinai}. Moreover sometimes it becomes
necessary to consider the wave function as a random process even in
closed systems (for example, when a classical analogue of the
quantum system has the features of the dynamical chaos)
\cite{avt1}-\cite{avt3}.

To be able to describe the cases mentioned above, in the present
paper we propose a radically new scheme of derivation of the
evolution equation for a nonrelativistic quantum system,
interacting with the thermostat (in particular with the physical
vacuum), 
with the wave function represented by a complex-valued random
process. The main idea of the new representation may be described
as follows: a potential energy of the system "quantum object +
thermostat" is assumed to be a random function. This is the case
where the Schr\"{o}dinger equation may be used only locally on
small time intervals and may provide good phenomenological models
for some problems, which are solvable within the framework of the
multiparticle Schr\"{o}dinger equation.

In the majority of cases the main tool for investigating the
particular problems is the perturbation theory, which sometimes
fails to provide an adequate description of a real physical
phenomenon. In the present paper the influence of the thermostat on
an elementary process is considered nonperturbatively within the
framework of randomly wandering one-dimensional quantum harmonic
oscillator model. Average transition probabilities for the
parametric oscillator are calculated, exact representations are
found for both widening and shift (analogous to the Lamb shift) of
the energy level of the oscillator submerged into the thermostat
(vacuum) as well as the entropy of an individual quantum state is
calculated.


\section{Description of the problem}
\label{sec-2}

We shall consider the closed system "quantum object + thermostat"
within the framework of a complex-valued probability process
representation. The realization of the process is a wave functional
$
\Psi_{stc}(x,t\vert \{ \vec \zeta \} ), $ defined on $ L_2\left
(R^1\otimes R_{\{ \vec \zeta \} }\right ) $ space (extended space),
where $ \vec \zeta (t) $ denotes a many-dimensional complex-valued
random process. Time evolution of the wave functional is governed
by the equation
\begin{equation}
i\partial_{t}\Psi_{stc}=\hat H \Psi_{stc},
\label{001}
\end{equation}
where one-dimensional Hamilton function $ \hat H $ is assumed to be
quadratic over the space variable
\begin{equation}
\hat H = -\frac {1}{2}\frac {\partial^2}{\partial x^2}+
\frac {1}{2}\Omega^2(t)x^2-F(t)x,
\label{002}
\end{equation}
and the functions $ \Omega^2(t) $ and $ F(t) $ are random functions
of time. Let them have the form
\begin{eqnarray}
\Omega^2(t) = \Omega_0^2(t) + \sqrt{2\epsilon_1p_1}f_1(t)\Theta(t-t_1),
\nonumber
\\
F(t) = F_0(t) + \sqrt{2\epsilon_2p_2}f_2(t)\Theta(t-t_2),
\label{003}
\end{eqnarray}
where $ \Omega_0^2(t) $ and $ F_0(t) $ are regular (nonrandom)
functions and $ f_1(t), f_2(t) $ are independent Gaussian random
processes with the zero mean and $ \delta - $shaped correlators
\begin{equation}
<f_i(t)f_j(t')> = \delta_{ij}\delta (t-t'), \quad i,j = 1,2.
\label{004}
\end{equation}
Constants $ \epsilon_i,\ i=1,2 $ control the power of forces $ f_i(t),\ t=1,2. $ 
Functions $ p_i(t),\ i=1,2 $ are assumed
nonnegative: $ p_1,p_2\ge 0. $ Step-function $ \Theta (x) $ is
defined by
\begin{equation}
\Theta (x)=\left \{
\begin{array}{ll}
1&x>0\\
0&x<0.
\end{array} \right.
\label{stupenka}
\end{equation}
Let us assume that the following asymptotic conditions hold
\begin{equation}
\Omega_0(t)\mathrel{\mathop{\longrightarrow }\limits_{t\rightarrow
\pm \infty }}\Omega_{\left (
in\atop out\right )},\quad
F_0(t)\mathrel{\mathop{\longrightarrow}\limits_{t\rightarrow
\pm \infty }} 0,\quad
p_i(t)\mathrel{\mathop{\longrightarrow}\limits_{t\rightarrow
\pm \infty }} 0,\ i=1,2.
\label{005}
\end{equation}
which guarantee that the autonomous states $ \phi_n^{in}(x,t) $
exist as $t\rightarrow -\infty $
\begin{eqnarray}
\phi_n^{in}(x,t) = e^{-i(n+1/2)\Omega_{in}t}\phi_n^{in}(x),
\quad \quad \quad \quad
\nonumber
\\
\phi_n^{in}(x) = \left (\frac {1}{2^nn!}\sqrt{\frac {\Omega_{in}}{\pi}}
\right )^{1/2}e^{-\Omega_{in}x^2/2}H_n(\sqrt{\Omega_{in}}x),
\label{006}
\end{eqnarray}
where $ \phi_n^{in}(x) $ is the wave function of a stationary
oscillator and $ H_n(x) $ is the Hermitian polynomial. It also
follows from (\ref{005}) that autonomous states $ \phi_n^{out}(x,t), $ 
which are obtained from (\ref{006}) by a substitution of $
\Omega_{in} $ by $
\Omega_{out}$, exist in the limit $ t\rightarrow +\infty $ as well. $ \Theta
$-functions in (\ref{003}) reflect the fact that the random
processes $ f_1(t) $ and $ f_2(t) $ are activated at the moments $
t_1 $ and $ t_2 $, respectively. If necessary, the functions $ p_1
$ and $ p_2 $ may be chosen having the form which prevents the
jumps of $\Omega $ and $ F $ when the noise is activating. The
moments $ t_1 $ and $ t_2 $ are assumed to be finite to make the
following inference correct. The aim of the paper is to find the
average probabilities $ W_{nm} $ of transitions from the initial
stationary states $ \phi_n^{in}(x,t) $ 
to the final ones $ \phi_m^{out}(x,t) $ when
the evolution is governed by Hamilton function (\ref{002}). Exact
mathematical definition of $ W_{nm} $ will be given in the next
section.


\section{Formal expressions for the wave functional and
transition probabilities}
\label{sec-3}

\begin{prop}
The formal solution of the problem (\ref{001})-(\ref{002}) may be written
down explicitly for arbitrary $ \Omega^2(t) $  and $ F(t) .$ It has the
following form
\begin{equation}
\Psi_{stc}(x,t\vert \{ \vec \zeta \} )=\frac {1}{\sqrt{r}}exp{\left \{i\left [\dot \eta (x-
\eta )+\frac {\dot r}{2r}(x-\eta )^2+\sigma \right ]\right \}}
\chi \left (\frac {x-\eta }{r},\tau \right ),
\label{101}
\end{equation}
where function $ \chi (y,\tau ) $ satisfies the Schr\"{o}dinger
equation for a harmonic oscillator with the constant frequency $
\Omega_{in} $
$$
i\frac {\partial \chi }{\partial \tau }=-\frac 1 2
\frac {\partial^2\chi }{\partial^2y}+\frac {\Omega_{in}^2y^2}{2}\chi ,
$$
function $ \eta (t) $ is a solution of the classical equation of
motion for the oscillator with the frequency $ \Omega(t), $
subjected to the external force $ F(t): $
\begin{equation}
\ddot \eta +\Omega^2(t)\eta =F(t),\quad \eta (-\infty )=
\dot \eta (-\infty )=0,
\label{102}
\end{equation}
$ \sigma (t) $ is a classical action, corresponding to the solution
$ \eta (t) $
\begin{equation}
\sigma (t)=\int_{-\infty }^t\left [\frac 1 2\dot \eta^2-
\frac 1 2\Omega^2\eta^2+F\eta \right ]dt',
\label{103}
\end{equation}
and $ r(t) $ and $ \tau (t) $ are expressed in terms of the
solution
$
\xi (t) $ of the homogeneous equation, corresponding to (\ref{102})
\begin{equation}
\ddot \xi +\Omega^2(t)\xi =0,\quad
\xi (t)\mathrel{\mathop{\sim }\limits_{t\rightarrow -\infty }}
e^{i\Omega_{in}t}
\label{104}
\end{equation}
as
$$
\xi (t)=r(t)e^{\gamma (t)},\quad r(t)=\vert \xi (t)\vert,\quad
\tau =\gamma (t)/\Omega_{in}.
$$
There was introduced a special designation $ \vec \zeta =(\xi ,\eta
) $ for the set of functions $ \xi (t) $ and $ \eta (t) $ from
(\ref{101}).
\end{prop}
\prf
The proof is based on the substitutions first used in \cite{Baz'}
and may be performed by the explicit verification under the only
suggestion that all the executed manipulations are legal.
\eprf

The set of solutions of type (\ref{101}) which is important for the
following considerations in this paper is obtained from (\ref{101})
after the substitution of $ \chi (y,\tau ) $ by $
\phi_n^{in}(y,\tau ) $ from (\ref{006}). It is thus defined as
\begin{equation}
\Psi_{stc}^{(n)}(x,t\vert \{ \vec \zeta \} )=
\frac {1}{\sqrt{r}}exp{\left \{i\left [\dot \eta (x-
\eta )+\frac {\dot r}{2r}(x-\eta )^2+\sigma \right ]\right \}}
\phi_n^{in}(\frac {x-\eta }{r},\tau ),
\label{105}
\end{equation}
$$
 n=1,2,... .
$$
The main properties of the set of functionals (\ref{105}), which
are important in what follows are
\begin{enumerate}
\item For any $ n $ the functional
$ \Psi_{stc}^{(n)}(x,t\vert \{ \vec \zeta \} ) $ reduces to the
autonomous state $ \phi_n^{in}(x,t) $ in the limit $ t\rightarrow
-\infty
.
$
\item For any fixed $ \vec \zeta $ elements of the set (\ref{105}) are mutually
orthogonal in the sense of $ L_2(R^1) $, space of square-integrable
functions:
\begin{equation}
\stackrel{\infty }{\mathrel{\mathop{\int }\limits_{-\infty }}}
\Psi_{stc}^{(n)}(x,t\vert \{ \vec \zeta \} )
\overline{\Psi_{stc}^{(m)}(x,t\vert \{ \vec \zeta \} )}\
dx=\delta_{nm},
\label{orthog}
\end{equation}
\end{enumerate}
where a bar denotes the complex conjugation procedure and $
\delta_{nm}=1, $ for $ n=m $ and $ \delta_{nm}=0 $ for $ n\neq m.
$
\begin{defin}
Average probabilities $ W_{nm} $ of transitions from the states $
\Psi_{stc}^{(n)}(x,t\vert \{ \vec \zeta \} ) $ to the stationary
ones $ \phi_m^{out}(x,t) $ in the limit $ t\rightarrow +\infty $
are defined by
\begin{equation}
\Psi_{stc}^{(n)}(x,t\vert \{ \vec \zeta \} )=
\sum_{m=0}^\infty c_{nm}(t\vert \{ \vec \zeta \} )
\phi_m^{out}(x,t),
\label{razlogh}
\end{equation}
\begin{equation}
W_{nm} =\lim_{t\rightarrow +\infty }\left <\vert c_{nm}\vert^2
\right >,
\label{106}
\end{equation}
where the symbol $ <..> $ denotes the procedure of averaging with
respect to $ f_1 $ and $ f_2. $
\end{defin}
\begin{defin}
The generating function $ I_{stc}(z_1,z_2,t\vert \{ \vec \zeta \} )
$ for coefficients $ c_{nm} $ is defined by the expression
\begin{equation}
I_{stc}(z_1,z_2,t\vert \{ \vec \zeta \} )=
\int_{-\infty }^{+\infty }dx \overline{\Psi_{out}(\bar z_1,x,t)}
\Psi_{stc}(z_2,x,t\vert \{ \vec \zeta \} ),
\label{gener}
\end{equation}
where
$$
\Psi_{stc}(z,x,t\vert \{ \vec \zeta \} )=\sum_{n=0}^\infty \frac {z^n}{\sqrt{n!}}
\Psi_{stc}^{(n)}(x,t\vert \{ \vec \zeta \} ),
$$
$$
\Psi_{out}(z,x,t)=\sum_{n=0}^\infty \frac {z^n}{\sqrt{n!}}
\phi_{n}^{(out)}(x,t),
$$
so that
\begin{equation}
c_{nm}(t\vert f_1,f_2)=\frac {\partial^{n+m}I_{stc}}
{\partial z_1^n\partial z_2^m}\Bigg \vert_{z_1=z_2=0}.
\label{coeff}
\end{equation}
\end{defin}
\begin{prop}
Explicit expression for the generating function (\ref{gener}) is
given by
\begin{equation}
I_{stc}(z_1,z_2,t\vert \{ \vec \zeta \} )=\left (\frac {2\sqrt{\Omega_{in}\Omega_{out}}}{K\xi }
\right )^{1/2}exp\{ Az_1^2+Bz_2^2+Cz_1z_2+Dz_1+Lz_2+M\}
\label{107}
\end{equation}
where the following notations are made
$$
A=\frac 1 2e^{2i\Omega_{out}t}\left (\frac {2\Omega_{out}}{K}-1\right ),
\quad B=\frac 1 2\left (\frac {2\Omega_{in}}{K\xi^2}-
e^{-2i\gamma }\right ),
$$
$$
C=\frac {2\sqrt{\Omega_{in}\Omega_{out}}}{K\xi }e^{i\Omega_{out}t}, \quad
L=-\frac {\sqrt{2\Omega_{in}}}{K\xi }(\Omega_{out}\eta -i\dot \eta ),
$$
$$
D=\sqrt{2\Omega_{out}}e^{i\Omega_{out}t}\left [\left (1-
\frac {\Omega_{out}}{K}\right )\eta +\frac {i\dot \eta }{K}\right ]
$$
$$
M=\frac {\Omega_{out}}{2}\left (\frac {\Omega_{out}}{K}-1\right )\eta^2
-\frac {1}{2K}\dot \eta^2-\frac {i\Omega_{out}}{K}\eta \dot \eta +
i\left ( \frac 1 2\Omega_{out}t+\sigma \right ),
$$
$$
K=-i\frac {\dot \xi }{\xi }+\Omega_{out }.
$$
\end{prop}
\prf The proof is carried out by the direct summation of the series over
Hermitian polynomials followed by the calculation of the Gaussian
integral.
\eprf

Expanding the expression (\ref{107}) in $ z_1 $ and $ z_2 $ powers,
we obtain the coefficients $ c_{nm}. $ The first several of them
are given below
\begin{eqnarray}
c_{00}=\left (\frac {2\sqrt{\Omega_{in}\Omega_{out}}}{K\xi }
\right )^{1/2}e^M,\quad c_{01}=Dc_{00},\quad c_{10}=Lc_{00},
\quad \quad \quad \quad
\label{108}
\\
c_{11}=(C+DL)c_{00},\quad c_{20}=\sqrt{2}\left (B+
\frac {L^2}{2}\right )c_{00},\quad c_{02}=\sqrt{2}
\left (A+\frac {D^2}{2}\right )c_{00}.
\nonumber
\end{eqnarray}
Given formal expressions for the objects to be averaged it is
necessary to reduce the averaging procedure to a form convenient
for the subsequent analytical or numerical treatment. The following
sections are devoted to the solution of this problem in different
situations, which may occur when considering the general problem
 (\ref{001})-(\ref{005}).


\section{Average transition probabilities in case of $ \epsilon_1=0 $}
\label{sec-4}

If $ \epsilon_1=0 $ the function $ \xi (t) $ from (\ref{104}) is
nonrandom. We denote it as $ \xi_0(t), $ and get
\begin{equation}
\ddot \xi_0 +\Omega_0^2(t)\xi_0 =0,\quad \xi_0(t)
\mathrel{\mathop{\sim }\limits_{t\rightarrow -\infty }}
e^{i\Omega_{in}t},
\label{201}
\end{equation}
$$
\xi_0(t)=r_0(t)e^{\gamma_0(t)}=\xi_{01}(t)+i\xi_{02}(t).
$$
We also introduce the designation $ \eta_0(t) $ for a function
satisfying the equation
\begin{equation}
\ddot \eta_0 +\Omega_0^2(t)\eta_0 =F_0(t),\quad \eta_0(-\infty )=
\dot \eta_0(-\infty )=0
\label{202}
\end{equation}
\begin{theor}
For any quantity $ G(\eta (t),\dot \eta (t)) $ local with respect
to $ \eta (t) $ and $ \dot \eta (t) $ (such are the coefficients $
c_{nm} $ in case of $ \epsilon_1=0 $) the averaging formula has the
following form
\begin{equation}
\left <G(\eta (t),\dot \eta (t))\right >=
\stackrel{+\infty }{\mathrel{\mathop{\int \int }\limits_{-\infty }}}
 dx_1 dx_2 \ G(x_1,x_2)P_1(x_1,x_2,t\vert
\eta_0(t_2),\dot \eta_0(t_2),t_2), \quad t>t_2.
\label{theor1}
\end{equation}
where
\begin{equation}
P_1=\frac {(4b_1b_3-b_2^2)^{-1/2}}{2\pi \Omega_{in} }
exp\left \{ -\frac {b_3y_1^2+b_1y_2^2-
b_2y_1y_2}{4b_1b_3-b_2^2}\right \},
\label{207}
\end{equation}
and
$$
b_1(t)=\frac {\epsilon_2}{\Omega_{in}^2}
\stackrel{t}{\mathrel{\mathop{\int }\limits_{t_2}}}
p_2(t')\xi_{01}^2(t') dt',
\quad b_3(t)=\frac {\epsilon_2}{\Omega_{in}^2}
\stackrel{t}{\mathrel{\mathop{\int }\limits_{t_2}}}
p_2(t')
\left (\dot \xi_{01}(t')\right )^2 dt',
$$
$$
b_2(t)=\frac {2\epsilon_2}{\Omega_{in}^2}
\stackrel{t}{\mathrel{\mathop{\int }\limits_{t_2}}}
p_2(t')\xi_{01}(t')\dot \xi_{01}(t')dt',
$$
$$
\left \{
\begin{array}{l}
y_1=-\left [\dot \xi_{01}(t)\Big (x_1-\eta_0(t)\Big )-\xi_{01}(t)
\Big (x_2-\dot \eta_0(t)\Big )\right ]/\Omega_{in},\\ \\
y_2=\left [\dot \xi_{02}(t)\Big (x_1-\eta_0(t)\Big )-\xi_{02}(t)
\Big (x_2-\dot \eta_0(t)\Big )
\right ]/\Omega_{in};\\
\end{array}\right.
$$
\end{theor}
\prf
It is obvious that the function $ \eta (t) $ is nonrandom in the
time interval $ t<t_2: $ $ \eta (t)=\eta_0(t), $ while in the
interval $ t>t_2 $ it represents a random process with the
evolution governed by the equation
\begin{equation}
\ddot \eta +\Omega_0^2(t)\eta =F_0(t)+\sqrt{2\epsilon_2p_2}f_2(t).
\label{203}
\end{equation}
The initial condition for (\ref{203}) is defined by the requirement
for the trajectory and its first derivative that they be continuous
at the moment $ t_2, $ i.e. $ \eta (t_2)=\eta_0(t_2),\
\dot \eta (t_2)=\dot \eta_0(t_2). $
To make the analysis of equation (\ref{203}), containing random
processes, correct it is convenient to rewrite it as the set of two
first order differential equations with respect to the quantities
 $ x_1=\eta,\ x_2=\dot \eta  $
\begin{equation}
\left\{
\begin{array}{l}
\dot x_1=x_2,\\
\dot x_2=F_0-\Omega_0^2x_1+\sqrt{2\epsilon_2p_2}f_2,\end{array}\right. \qquad
\left\{
\begin{array}{l}
x_1(t_2)=\eta_0(t_2),\\
x_2(t_2)=\dot \eta_0(t_2),\end{array}\right.
\label{204}
\end{equation}
Equations (\ref{204}) are naturally interpreted as SDE for the
random processes $ x_1(t) $ and $ x_2(t). $ Proceeding from them it
is not difficult to write down the Fokker-Planck equation for the
conditional probability density
$$
P_1(x_1,x_2,t\vert x_{10},x_{20},t_0)=\Big <\delta \big (x_1(t)-
x_1\big )\delta \big (x_2(t)-x_2\big )\Big >
\bigg \vert_{\begin{array}{l}
x_1(t_0)=x_{10}
\\
x_2(t_0)=x_{20},
\end{array}}
$$
describing the probability that trajectory $ \big
(x_1(t),x_2(t)\big ) $ finds itself in the vicinity of the point $
\big (x_1,x_2\big ) $ at the moment $ t, $ having started from the
point $ \big (x_{10},x_{20}\big ) $ at the moment $ t_0. $ It can
be shown that the equation for $ P_1 $ has the form (see
\cite{Gred} or \cite{Gard})
\begin{equation}
\frac {\partial P_1}{\partial t}=-x_2\frac {\partial P_1}{\partial x_1}-
(F_0-\Omega_0^2x_1)\frac {\partial P_1}{\partial x_2}+\epsilon_2p_2
\frac {\partial^2 P_1}{\partial x_2^2}.
\label{205}
\end{equation}
The solution of (\ref{205}) must be integrable and satisfy the
obvious initial condition
\begin{equation}
P_1\Big \vert_{t=t_0}=\delta (x_1-x_{10})\delta(x_2-x_{20})
\label{206}
\end{equation}
As the position of the trajectory at the moment $ t_2 $ is known,
it is natural to set $ t_0=t_2,\ x_{10}=\eta_0(t_2),\ x_{20}=\dot
\eta_0(t_2). $ The integrable solution of (\ref{205}), satisfying
(\ref{206}), may be expressed in terms of the functions $
\xi_{01}(t),\
\xi_{02}(t),\
\eta_0(t). $ It is easy to test that it has
the form (\ref{207}) by the explicit verification. The theorem is
proved.
\eprf

To find probabilities $ W_{nm} $ in this case we calculate the
Gaussian integrals and then compute their limiting values at $
t\rightarrow +\infty . $


\section{Average transition probabilities in case of $ \epsilon_2=0 $}
\label{sec-5}

In the case of $ \epsilon_2=0 $ the time axis is broken into two
parts. At $ t<t_1 $ functions $ \eta (t) $ and $ \xi (t) $ are
nonrandom:
$
\eta (t)=\eta_0(t),\ \xi (t)=\xi_0(t). $ At $ t>t_1 $ both $ \eta
(t) $'s and $ \xi (t) $'s trajectories become random.
\begin{theor}
In the case of $ \epsilon_2=0, $ at $ t>t_1 $ the set of equations
(\ref{102}), (\ref{104}) gives rise to the set of SDE, describing
the evolution of four random processes
$$
\vec u(t)\equiv \big (u_1(t),u_2(t),u_3(t),u_4(t)\big )\equiv
\left (\eta (t),\dot \eta (t),Re\left (\frac {\dot \xi (t)}
{\xi (t)}\right ),
Im\left (\frac {\dot \xi (t)}{\xi (t)}\right )\right )
$$
with the joint probability distribution function $ P_2(\vec
u,t\vert
\vec u_0,t_1),\ t>t_1, $ satisfying the Fokker-Planck equation
\begin{equation}
\frac {\partial P_2}{\partial t}=\hat L_2P_2,
\label{305}
\end{equation}
$$
\hat L_2\left (\vec u\right )\equiv
-\sum_{i=1}^4
K_i\frac {\partial }{\partial u_i}+
\epsilon_1p_1u_1^2\frac {\partial^2}{\partial u_2^2}+
\epsilon_1p_1\frac {\partial^2}{\partial u_3^2}+
2\epsilon_1p_1u_1\frac {\partial^2}{\partial u_2\partial u_3}
+4u_3,
$$
$$
K_1=u_2,\quad K_2=F_0-\Omega_0^2u_1,\quad K_3=u_4^2-u_3^2-
\Omega_0^2,\quad K_4=-2u_3u_4.
$$
and the initial condition
$$
P_2\big \vert_{t=t_1}=\delta \left (\vec u-\vec u_0\right )\equiv
\prod_{i=1}^4\delta (u_i-u_{0i}).
$$
\label{theor-4.1}
\end{theor}
\prf
In the case under consideration the equation (\ref{102}) is
transformed to a set of SDE in the same way as it has been done in
the previous section, namely, by introducing the quantities $
u_1=\eta,\ u_2=\dot \eta $ which reduce (\ref{102}) to (\ref{204})
with the only distinction in the initial condition: $
u_1(t_1)=\eta_0(t_1),\ u_2(t_1)=\dot
\eta_0(t_1). $

The equation (\ref{104}) may be reduced to a nonlinear first order
differential equation by the substitution
\begin{equation}
\xi (t)=
\left \{
\begin{array}{ll}
\xi_0(t),&t<t_1\\
\xi_0(t_1)\exp{\left \{
\stackrel{t}{\mathrel{\mathop{\int }\limits_{t_1}}}
\Phi(t')dt'\right \}},&t>t_1,
\end{array}
\right.
\label{302}
\end{equation}
which gives upon being applied to (\ref{104}) the following SDE for
$
\Phi (t) $ in the interval $ t_1<t<\infty $
\begin{equation}
\dot \Phi (t)+\Phi^2(t)+\Omega_0^2(t)+\sqrt{2\epsilon_1p_1}f_1=0,
\quad \Phi (t_1)=\dot \xi_0(t_1)/\xi_0(t_1).
\label{303}
\end{equation}
where the second equation expresses a condition which guarantees
continuity of the function $ \xi (t) $ and its first derivative at
$ t=t_1. $ The function $ \Phi (t) $ is a complex-valued random
process due to the initial condition. As a result the SDE
(\ref{303}) is equivalent to a set of two SDE for real-valued
random processes. Namely, introducing real and imaginary parts of $
\Phi (t) $
$$
\Phi (t) =u_3 (t) +iu_4 (t),
$$
we finally obtain the following set of SDE for the components of
the random vector process $ \vec u $
\begin{equation}
\left \{
\begin{array}{l}
\dot u_1=u_2,\\
\dot u_2=F_0-\Omega_0^2u_1-\sqrt{2\epsilon_1p_1}u_1f_1,\\
\dot u_3=-u_3^2+u_4^2-\Omega_0^2(t)-\sqrt{2\epsilon_1p_1}f_1(t),\\
\dot u_4=-2u_3u_4,\\
\end{array}
\right. \
\left \{
\begin{array}{lll}
u_1(t_1)=\eta_0(t_1),\\
u_2(t_1)=\dot \eta_0(t_1),\\
u_3(t_1)=Re\left (\dot \xi_0(t_1)/\xi_0(t_1)\right ),\\
u_4(t_1)=Im\left (\dot \xi_0(t_1)/\xi_0(t_1)\right ).\\
\end{array} \right.
\label{304}
\end{equation}
The pairs of random processes $ (u_1,u_2) $ and $ (u_3,u_4) $ are
not independent, because their evolution is influenced by the
common random force $ f_1(t). $ This means that the joint
probability distribution
$$
P_2\big (\vec u,t\vert \vec u_0,t_1\big )=
\left <\prod_{i=1}^4\delta \big (u_i(t)-u_i\big )\right >
\Biggl \vert_{\begin{array}{c} \\
\vec u(t_1)=\vec u_0
\end{array}}
$$
$$
\vec u_0=\left (\eta_0(t_1),\dot \eta_0(t_1),
Re\left (\dot \xi_0(t_1)/\xi_0(t_1)\right ),
Im\left (\dot \xi_0(t_1)/\xi_0(t_1)\right )\right )
$$
is a nonfactorable function. Proceeding from the known evolution
equations (\ref{304}), we obtain by the standard method the
Fokker-Planck equation for $ P_2 $ (see \cite{Gred} or
\cite{Gard}), which has the form (\ref{305}). The theorem is
proved.
\eprf

Given $ P_2, $ one can average any quantity $ G\left (\vec
u(t)\right ) $ which is local with respect to $ \vec u(t): $
\begin{equation}
\left <G\left (\vec u(t)\right )\right >=
\int d\vec u P_2\left (\vec u,t\vert
\vec u_0,t_1\right )G\left (\vec u\right ),\quad d\vec u=du_1du_2du_3du_4.
\label{306}
\end{equation}
But this formula fails to give a result if it is used for averaging
the objects containing the coefficients $ c_{nm} $ from
(\ref{coeff}), which are nonlocal with respect to $ \vec u. $ There
does not exist a general approach to calculating the average value
of any quantity nonlocal with respect to a random process. But it
is known that for some types of such objects the averaging
procedure may be reduced to finding a fundamental solution of some
parabolic partial differential equation and its subsequent weighted
integration. The description of the simplest case of this kind is
given in \cite{Kac}. It is not difficult to generalize the result
obtained in \cite{Kac}, and therefore the formulas
(\ref{1003})-(\ref{1004}) can be derived (see Appendix). Using
(\ref{1003})-(\ref{1004}), we have the following proposition.
\begin{prop}
If the components of the random vector process $ \vec u $ satisfy
the set of SDE (\ref{304}), then the averaging procedure can be
represented as
\begin{equation}
\left <\exp\left \{-\int_{t_1}^tV_1\left (\vec u(\tau ),\vec u(t),t\right )d\tau
-V_2\left (\vec u(t)\right )\right \} \right >=
\int d\vec u e^{-V_2\left (\vec u\right )}Q\left (\vec u,\vec u,t\right ),
\label{307}
\end{equation}
where the function $ Q\left (\vec u,\vec u ',t\right ) $ is a
solution of the problem
\begin{equation}
\begin{array}{l}
\displaystyle
\frac {\partial Q}{\partial t}=\left (\hat L_2\left (\vec u\right )-
V_1\left (\vec u,\vec u ',t\right )\right )Q,\qquad \quad
\\ \\
Q\left (\vec u,\vec u ',t\right )
\mathrel{\mathop{\longrightarrow }\limits_{t\rightarrow t_1}}
\delta \left (\vec u-\vec u_0\right ),
\quad
 Q\left (\vec u,\vec u ',t\right )
\mathrel{\mathop{\longrightarrow }\limits_{\vert \vert \vec u\vert
\vert \rightarrow \infty }}0,
\end{array}
\label{308}
\end{equation}
where $ \vert \vert \cdot \vert \vert $ is a norm in $ R^4. $
\label{prop-3}
\end{prop}

It is not difficult to show that when $ V_1=0, $ the formula
(\ref{307}) transforms into (\ref{306}) with the substitution $
G=\exp{\{-V_2\}} $, because in this case the equation (\ref{308})
for $ Q $ transforms into the Fokker-Planck equation (\ref{305})
for $ P_2 $.

Using the proposition \ref{sec-5}.\ref{prop-3}, we obtain the
representation for the average values $ \left <\vert
c_{nm}\vert^2\right >, $ which are equal to the probabilities $
W_{nm} $ in the limit $ t\rightarrow +\infty .$ The explicit
expressions for the first four of them are presented here $
(n=0,1;\ m=0,1): $
\begin{equation}
\left <\vert c_{nm}\vert^2\right >=
\int d\vec u H_{nm}\left (\vec u\right )Q_{nm}\left (\vec u,t\right ),
\label{309}
\end{equation}
where
\begin{eqnarray}
H_{00}\left (\vec u\right )&=&\frac {2\sqrt{\Omega_{in}\Omega_{out}}}
{\vert \xi_0(t_1)K\vert }\exp \left \{
\frac {\Omega_{out}}{2}\left (\frac {2\Omega_{out}(u_4+\Omega_{out})}
{\vert K\vert^2}-2\right )u_1^2- \right.
\nonumber
\\
& &
\left.
-\frac {u_4+\Omega_{out}}
{\vert K\vert^2}u_2^2+\frac {2\Omega_{out}u_3}{\vert K\vert^2}u_1u_2
\right \},
\nonumber
\\
H_{01}\left (\vec u\right )&=&\frac {2\Omega_{out}}{\vert K\vert^2}
\left ( (u_2-u_1u_3)^2+u_1^2u_4^2\right )H_{00}\left (\vec u\right ),
\nonumber
\\
H_{10}\left (\vec u\right )&=&\frac {2\Omega_{in}}
{\vert \xi_0(t_1)K\vert^2}\left (\Omega_{out}^2u_1^2+u_2^2\right )
H_{00}\left (\vec u\right ),
\label{310}
\\
H_{11}\left (\vec u\right )&=&\frac {4\Omega_{in}\Omega_{out}}
{\vert \xi_0(t_1)K^2\vert^2}\Big \vert K-
(\Omega_{out}u_1-iu_2)(u_1u_4+i(u_2-u_1u_3))\Big \vert^2
H_{00}\left (\vec u\right )
\nonumber
\end{eqnarray}
$$
K=\Omega_{out}+u_4-iu_3,
$$
and functions $ Q_{nm}\left (\vec u,t\right ) $ satisfy the
equations
\begin{equation}
\begin{array}{l}
\displaystyle
\frac {\partial Q_{nm}}{\partial t}=\hat L_2Q_{nm}-V_{nm}Q_{nm},
\\ \\
Q_{nm}\left (\vec u,t\right )
\mathrel{\mathop{\longrightarrow }\limits_{t\rightarrow t_1}}
\delta \left (\vec u-\vec u_0\right ),
\quad
 Q_{nm}\left (\vec u,t\right )
\mathrel{\mathop{\longrightarrow }\limits_{\vert \vert \vec u\vert
\vert \rightarrow \infty }}0,
\end{array}
\label{311}
\end{equation}
where
$$
V_{nm}=p_{nm}u_3,\quad p_{00}=p_{01}=1,\quad p_{10}=p_{11}=3
$$
To obtain $ W_{nm} $ it is necessary to proceed in (\ref{309}) to
the limit $ t\rightarrow +\infty . $ The representation
(\ref{309})-(\ref{311}) is exact and free from any simplifying
assumptions. Given a specific realization of $ p_1(t), $
(\ref{309})-(\ref{311}) is used as a basis for numerical
calculations of the probabilities $ W_{nm}. $

More simple representation may be obtained under some additional
assumptions. First we introduce some useful designations. Namely,
we shall start with the expression for the solution of equation
(\ref{201}) applicable at $ t\rightarrow +\infty : $
$$
\xi_0(t)=C_1e^{i\Omega_{out}t}+C_2e^{-i\Omega_{out}t},\quad
C_1=\vert C_1\vert e^{i\delta_1},\ C_2=\vert C_2\vert e^{i\delta_2},
$$
then we shall write down also a corresponding representation for
the solution of the equation (\ref{202}):
$$
\eta_0(t)=\frac 1 {\sqrt{2\Omega_{in}}}\left (\xi_0d^*+\xi_0^*d\right ),
\quad d(t)=\frac i {\sqrt{2\Omega_{in}}}
\stackrel{t}{\mathrel{\mathop{\int }\limits_{-\infty }}}
\xi_0(t')F_0(t')dt'.
$$
The following quantities, defined on the basis of the above
formulas, will be included into the final expressions for the
required probabilities
\begin{equation}
\rho =\left \vert \frac {C_2}{C_1}\right \vert^2,\quad
d=\lim_{t\rightarrow +\infty }d(t)=\sqrt {\nu }e^{i\beta },
\label{324}
\end{equation}
\begin{theor}
Let the function $ p_1(t) $ have the form
\begin{equation}
p_1(t)=
\left \{
\begin{array}{ll}
1&t_1<t<t_e,\\
0&t>t_e.
\end{array}
\right.
\label{312}
\end{equation}
and time $ t_e $ be large enough to guarantee that the solution $
Q_{nm}\left (\vec u,t\right ) $ of the equation (\ref{311}) at the
moment $ t_e $ may be replaced approximately by its stationary (at
$ t\to +\infty )$ limit. Then we can obtain the following
representation of the probabilities $ W_{nm} $
\begin{equation}
W_{nm}=\Omega_{in}^{p_{nm}}
\int d\xi_1d\xi_2d\xi_3
{\bar Q_{nm}^{st}(\xi_1,\xi_2,\xi_3)}
\bar H_{nm}\left (\xi_1, \xi_2, \xi_3\right ).
\label{325}
\end{equation}
where $ \bar Q_{nm}^{st} $ is a solution of the shortened
stationary equation
\begin{equation}
\left (\hat {\bar L}_2^{st}(\vec \xi )-p_{nm}\xi_3\right )
\bar Q_{nm}^{st}(\vec \xi )=0,
\label{322}
\end{equation}
$$
\hat {\bar L}_2^{st}\left (\vec \xi \right )\equiv
-\sum_{i=1}^3\bar K_i\frac {\partial }{\partial \xi_i}+
\epsilon_1\xi_1^2\frac {\partial^2 }{\partial \xi_2^2}+
\epsilon_1\frac {\partial^2 }{\partial \xi_3^2}+
2\epsilon_1\xi_1\frac {\partial^2 }{\partial \xi_2
\partial \xi_3}+2\xi_3.
$$
$$
\bar K_1=\xi_2,\quad \bar K_2=-\Omega_{out}^2\xi_1,\quad
\bar K_3=-(\xi_3^2+\Omega_{out}^2).
$$
Quantities $ \bar H_{nm} $ for $ n=0,1 $ and $ m=0 $ are given by
$$
\bar H_{00}(\xi_1, \xi_2, \xi_3)=\frac {2\sqrt {\Omega_{in}\Omega_{out}}}
{\vert \xi_0(t_1)\vert \sqrt {\Sigma (\xi_3)}}
\exp \left \{-\frac {\Omega_{out}\Omega_{in}^2}{\Sigma (\xi_3)}
[\xi_3(\xi_1+\mu_1)-\xi_2-\mu_2]^2\right \},
$$
$$
\bar H_{01}(\xi_1, \xi_2, \xi_3)=
\frac {2\Omega_{out}\Omega_{in}^2}{\Sigma (\xi_3)}
[\xi_2-\xi_1\xi_3-\xi_3\mu_1+\mu_2]^2\bar H_{00}(\xi_1, \xi_2, \xi_3),
$$
$$
\mu_1=-d_5+\sqrt {\frac {2\nu }{\Omega_{in}}}(d_1\cos \beta +d_2\sin \beta )
$$
\begin{equation}
\mu_2=-d_6+\sqrt {\frac {2\nu }{\Omega_{in}}}(d_3\cos \beta +d_4\sin \beta )
\label{326}
\end{equation}
$$
\Sigma (\xi_3)=\frac {\Omega_{in}\Omega_{out}}{(1-\rho )}
\left \{ \biggl [(d_1^2+d_2^2)(1+\rho )-2\sqrt {\rho }
\Bigl [(d_1^2-d_2^2)\cos \delta +2d_1d_2\sin
\delta \Bigr ]\biggr ]\xi_3^2 \right.+
$$
$$
\left.+2\biggl [-(d_2d_4+d_1d_3)(1+\rho )+2\sqrt {\rho }\Bigl [
(d_1d_4+d_2d_3)\sin \delta +
(d_1d_3-d_2d_4)\cos \delta \Bigr ]\biggr ]\xi_3+\right.
$$
$$
\left.+\biggl [(d_3^2+d_4^2)(1+\rho )+2\sqrt {\rho }
\Bigl [(d_4^2-d_3^2)\cos \delta -2d_3d_4\sin
\delta \Bigr ]\biggr ] \right \}.\qquad \quad
$$
where $ \delta =\delta_1+\delta_2. $ The ambiguity which arises in
finding the solution $ \bar Q_{nm}^{st} $ of the equation
(\ref{322}) is eliminated by the requirement of the correspondence
of $\bar Q_{nm}^{st} $ to the stationary limit of the solution $
Q_{nm} $ of the equation (\ref{311}).
\label{theor-3}
\end{theor}
\prf
The proof originates from the representation
(\ref{309})-(\ref{311}). Denote the solution of the problem
(\ref{311}) in the time interval
 $ t_1<t<t_e $ by $ Q_{nm}^<\left (\vec u,t\right ) $ and that in the
time interval $ t>t_e $ by $ Q_{nm}^>\left (\vec u,t\right ). $
Then $ Q_{nm}^>\left (\vec u,t\right ) $ satisfies the first order
differential equation
\begin{eqnarray}
\frac {\partial Q_{nm}^>}{\partial t}=-\sum_{i=1}^4
\frac {\partial (K_iQ_{nm}^>)}{\partial u_i}-V_{nm}Q_{nm}^>,
\quad
\nonumber
\\
\quad Q_{nm}^>\left (\vec u,t\right )
\mathrel{\mathop{\longrightarrow }\limits_{t\rightarrow t_e}}
Q_{nm}^<\left (\vec u,t_e\right ),
\qquad
 Q_{nm}^>\left (\vec u,t\right )
\mathrel{\mathop{\longrightarrow }\limits_{\vert \vert \vec u\vert
\vert \rightarrow \infty }}0.
\label{313}
\end{eqnarray}
Using the method of characteristics the problem (\ref{313}) can be
solved for an arbitrary initial condition. So that we obtain the
following expression containing the solutions of classical
equations of motion (\ref{201}) and (\ref{202}):
\begin{equation}
Q_{nm}^>\left (\vec u,t\right )=Q_{nm}^<\left (\vec \xi ,t_e\right )\cdot
\exp \left ((4-p_{nm})\int_{t_e}^t \  u_3\left (\vec \xi , \tau \right )d\tau \right ),
\label{314}
\end{equation}
Quantities $ \xi_i\left (\vec u,t\right ),\ i=1,..,4 $ in
(\ref{314}) are the first integrals of the characteristic set of
ordinary first order differential equations, corresponding to the
equation (\ref{312}). Their dependency on $ \vec u $ variables is
given implicitly by the relations
\begin{equation}
\begin{array}{ll}
u_1={\displaystyle \frac 1{\Omega_{in}}}
\big [e_{22}\xi_1+e_{21}\xi_2+h_1\big ],&
u_2={\displaystyle \frac 1{\Omega_{in}}}
\big [e_{12}\xi_1+e_{11}\xi_2+h_2\big ],
\\ \\
u_3={\displaystyle \frac {(e_{11}\xi_3+e_{12})(e_{21}\xi_3+e_{22})+e_{11}e_{21}
\xi_4^2}{(e_{21}\xi_3+e_{22})^2+e_{21}^2\xi_4^2}},&
u_4={\displaystyle \frac {\Omega_{in}\xi_4}
{(e_{21}\xi_3+e_{22})^2+e_{21}^2\xi_4^2}}.
\end{array}
\label{315}
\end{equation}
The definitions of functions and constants appearing in (\ref{315})
are given by
$$
h_1(t)=(d_2d_6-d_4d_5)\xi_{01}(t)+(d_3d_5-d_1d_6)\xi_{02}(t)+
\Omega_{in}\eta_0(t),\quad h_2(t)=\dot h_1(t); \quad
$$
\begin{equation}
\begin{array}{ll}
e_{21}(t)=d_1\xi_{02}(t)-d_2\xi_{01}(t),&
e_{12}(t)=\dot e_{22}(t),
\\
e_{22}(t)=d_4\xi_{01}(t)-d_3\xi_{02}(t),
& e_{11}(t)=\dot e_{21}(t);
\\
d_1=\xi_{01}(t_e),\ d_2=\xi_{02}(t_e),&
d_5=\eta_0(t_e),
\\
d_3=\dot \xi_{01}(t_e),\ d_4=\dot \xi_{02}(t_e),
&d_6=\dot \eta_0(t_e).
\end{array}
\label{316}
\end{equation}
The functions $ \xi_{01}(t) ,\ \xi_{02}(t) $ and $ \eta_0(t) $ are
defined in (\ref{201}) and (\ref{202}). By the explicit
verification it may be proved that the new variables satisfy the
initial conditions
\begin{equation}
\xi_i\left (\vec u,t_e\right )=u_i,\quad i=1,..,4,
\label{317}
\end{equation}
so that the solution (\ref{314}) satisfies the initial condition
defined in (\ref{313}).

Using (\ref{315}), the Jacobian
\begin{equation}
J\left (\vec \xi \right )\equiv \frac {\partial (u_1,..,u_4)}{\partial (\xi_1,..,\xi_4)}
\label{jacob}
\end{equation}
of transition to new variables is calculated. In case of $ \xi_4=0
$, which is important in what follows, it takes form
\begin{equation}
J\left (\vec \xi \right )\Big \vert_{\xi_4=0}=\frac {\Omega_{in}^4}{(e_{21}\xi_3+
e_{22})^4}
\label{318}
\end{equation}
Proceeding from the above we can rewrite the representation
(\ref{309}) as:
\begin{equation}
\left <\vert c_{nm}\vert^2\right >=
\int d\vec \xi \
J\left (\vec \xi \right )
H_{nm}\Bigl (\vec u(\vec \xi ,t)\Bigr )Q_{nm}^<\left (\vec \xi ,t_e\right )
\exp \left ((4-p_{nm})
\stackrel{t}{\mathrel{\mathop{\int }\limits_{t_e}}}
u_3\left (\vec \xi , \tau \right )d\tau \right ),
\label{319}
\end{equation}
where integration over $ \tau $ is carried out under the assumption
that $ \vec \xi $ is constant. Assuming that the moment $ t_e $ is
big enough the solution $ Q_{nm}^<\left (\vec \xi ,t_e\right ) $
can be replaced approximately by the stationary limit $
Q_{nm}^{st}\left (\vec \xi
\right ), $ which satisfies the equation
\begin{equation}
\left (\hat L_2^{st}-V_{nm}\right )Q_{nm}^{st}=0,\quad
\hat L_2^{st}=\lim_{t\rightarrow +\infty }\hat L_2
\label{320}
\end{equation}
Operator $ \hat L_2 $ is defined in (\ref{305}). If necessary an
exact value of $ t_e $, which will be sufficient for a legality of
the replacement mentioned above, may be estimated. It is obvious
that the stationary limit $ Q_{nm}^{st}(\vec u) $ of the function $
Q_{nm}(\vec u,t) $ depends on the initial condition for the latter.

The insertion of the stationary solution $ Q_{nm}^{st} $ into the
formulas make them more simple, because the stationary equation
allows a separation of variables. So substituting
\begin{equation}
Q_{nm}^{st}=\delta (\xi_4)\bar Q_{nm}^{st}(\xi_1,\xi_2,\xi_3),
\label{321}
\end{equation}
into (\ref{320}), we obtain the following (shortened) equation
determining the function $\bar Q_{nm}^{st} $
\begin{equation}
\left (\hat {\bar L}_2^{st}(\vec \xi )-p_{nm}\xi_3\right )
\bar Q_{nm}^{st}(\vec \xi )=0,
\label{322dop}
\end{equation}
$$
\hat {\bar L}_2^{st}\left (\vec \xi \right )\equiv
-\sum_{i=1}^3\bar K_i\frac {\partial }{\partial \xi_i}+
\epsilon_1\xi_1^2\frac {\partial^2 }{\partial \xi_2^2}+
\epsilon_1\frac {\partial^2 }{\partial \xi_3^2}+
2\epsilon_1\xi_1\frac {\partial^2 }{\partial \xi_2
\partial \xi_3}+2\xi_3.
$$
$$
\bar K_1=\xi_2,\quad \bar K_2=-\Omega_{out}^2\xi_1,\quad
\bar K_3=-(\xi_3^2+\Omega_{out}^2).
$$
Substituting (\ref{321}) into (\ref{319}) and using (\ref{318}), we
come to the formula
\begin{equation}
\left <\vert c_{nm}\vert^2\right >=\Omega_{in}^{p_{nm}}
\int d\xi_1d\xi_2d\xi_3
\frac {\bar Q_{nm}^{st}(\xi_1,\xi_2,\xi_3)}{(e_{21}\xi_3+e_{22})^{p_{nm}}}
H_{nm}\left (\vec u(\vec \xi ,t)\right )\bigl \vert_{\xi_4=0}.
\label{323}
\end{equation}
To find the average probabilities $ W_{nm} $ we proceed to the
limit $ t\rightarrow +\infty $ in (\ref{323}). This may be done for
any $ n $ and $ m. $ After implementing simple but wearisome
transformations the limiting values of (\ref{323}) were obtained
for the minimal $ n,\ m. $ They have the form
(\ref{325})-(\ref{326}). The theorem is proved.
\eprf

It is a real challenge to obtain the solution $ \bar Q_{nm}^{st} $,
corresponding to a given initial condition appearing in
(\ref{311}), without solving the equation on the whole interval $
t_1<t<\infty $. But this problem may be resolved in the following
particular case.
\begin{theor}
Let $ F_0(t)\equiv const=F_0, $ $ \Omega_0(t)\equiv const=\Omega_0,
$ then $ \bar Q_{nm}^{st} $ entering into the statement of the
theorem \ref{sec-5}.\ref{theor-3} is given by
\begin{equation}
\bar Q_{nm}^{st}(u_1,u_2,u_3)=C_{nm}\bar q_{nm}^{st}(u_1,u_2,u_3),
\label{add3.1}
\end{equation}
where $ \bar q_{nm}^{st}(u_1,u_2,u_3) $ is an arbitrary fixed
solution of (\ref{322dop}) decreasing on infinity, and the constant
$ C_{nm} $ is expressed in terms of this solution as follows
\begin{equation}
C_{nm}= {Y_{nm}^{st}(\vec u_0)}
\left \{\int du_1du_2du_3\
Y_{nm}^{st}(\vec u)\Big \vert_{u_4=0}\bar q_{nm}^{st}(u_1,u_2,u_3)\right \}^{-1}
\label{add3.44}
\end{equation}
where $ Y_{nm}^{st} $ is any solution of the conjugate equation
\begin{equation}
\left (\left (\hat {L}_2^{st}\right )^+(\vec u)-p_{nm}u_3\right )
Y_{nm}^{st}(\vec u)=0,
\label{add.adjoint}
\end{equation}
$$
\left (\hat {\bar L}_2^{st}\right )^+\left (\vec u\right )\equiv
\sum_{i=1}^4K_i^{st}\frac {\partial }{\partial u_i}+
\epsilon_1u_1^2\frac {\partial^2 }{\partial u_2^2}+
\epsilon_1\frac {\partial^2 }{\partial u_3^2}+
2\epsilon_1u_1\frac {\partial^2 }{\partial u_2
\partial u_3}.
$$
$$
 K_1^{st}=u_2,\quad K_2^{st}=-\Omega_{out}^2u_1,\quad
 K_3^{st}=u_4^2-u_3^2-\Omega_{out}^2),\quad K_4^{st}=-2u_3u_4.
$$
\label{theor-4.3}
\end{theor}
\prf
To prove the above statement we consider the quantity
\begin{equation}
I_{nm}^{(Y)}(t)\equiv \int d\vec u\ Y_{nm}(\vec u,t)Q_{nm}(\vec u,t),
\label{add3.2}
\end{equation}
where $ Q_{nm}(\vec u,t) $ is a solution of the problem (\ref{311})
and function $ Y_{nm}(\vec u,t) $ is assumed to be found proceeding
from the requirement that (\ref{add3.2}) does not depend on the
time variable, i.e. $ dI_{nm}^{(Y)}/dt=0. $ Integrating by parts we
can easily show that the requirement will be satisfied if the
function $ Y_{nm} $ is an arbitrary solution of the equation
conjugate to (\ref{311})
\begin{equation}
\frac {\partial Y_{nm}}{\partial t}=\left (-\hat L_2^++V_{nm}\right )Y_{nm},
\label{add3.3}
\end{equation}
$$
\hat L_2^+\left (\vec u\right )\equiv
\sum_{i=1}^4
K_i\frac {\partial }{\partial u_i}+
\epsilon_1p_1u_1^2\frac {\partial^2}{\partial u_2^2}+
\epsilon_1p_1\frac {\partial^2}{\partial u_3^2}+
2\epsilon_1p_1u_1\frac {\partial^2}{\partial u_2\partial u_3}.
$$
The only condition which is supposed to be imposed on $ Y_{nm} $ is
its sufficiently rapid decreasing with $ \vert \vert \vec u\vert
\vert \to \infty $. This guarantees that all the terms,
containing integration over infinitely remote surfaces, will
vanish.

It is worth mentioning that the quantity (\ref{add3.2}) does not
depend on time even when the functions $ \Omega_0(t) $ and $ F_0(t)
$ entering into the statement of the theorem are nonconstant. This
condition becomes important in the case when the constancy of the
quantity (\ref{add3.2}) is used for the restoration of information
about the initial condition, which is lost in proceeding to the
stationary limit in (\ref{311}).

If $ p_1(t)=p_2(t)\equiv 1, $ $ F_0(t)\equiv const=F_0, $ $
\Omega_0(t)\equiv const=\Omega_0 $ the operator $ \hat L_2^+ $
which is conjugate to the Fokker-Planck operator $ \hat L_2 $ does
not depend on time. In this case it may be the time independent
solution $ Y_{nm}^{st} $ of the equation (\ref{add3.3})
\begin{equation}
\left (-\hat L_2^++p_{nm}u_3\right )Y_{nm}^{st}=0.
\label{add3.5}
\end{equation}
to be substituted into (\ref{add3.2}) instead of $ Y_{nm}. $ Note
that in this case $ \hat L_2^+=\left (\hat L_2^{st}\right )^+, $ so
that the equation (\ref{add3.5}) is in fact a stationary equation
corresponding to (\ref{add3.3}). Having chosen such a function $
Y_{nm} $ one can equate values of the quantity (\ref{add3.2}) at $
t=t_1 $ and at $ t\to \infty : $
\begin{equation}
{Y_{nm}^{st}(\vec u_0)}={\int d\vec uY_{nm}^{st}(\vec u)
Q_{nm}^{st}(\vec u)}.
\label{add.eqv}
\end{equation}
Proceeding from an arbitrary solution $ q_{nm}^{st}(\vec u) $ of
the stationary equation (\ref{320}) with a sufficiently rapid
decreasing on infinity we obtain the required solution $
Q_{nm}^{st}(\vec u) $
\begin{equation}
Q_{nm}^{st}(\vec u)=C_{nm}q_{nm}^{st}(\vec u),
\label{add3.1dop}
\end{equation}
where the constant $ C_{nm} $ depends on the specific choice of $
q_{nm}^{st}(\vec u). $ Unfortunately, so far we do not possess a
strict proof of such representation, but it seems plausible in view
of the results obtained in exploring the more simple
one-dimensional equation. Substituting (\ref{add3.1dop}) into
(\ref{add3.2}), one readily obtains
\begin{equation}
C_{nm}= \displaystyle \frac {Y_{nm}^{st}(\vec u_0)}{\int d\vec u\
Y_{nm}^{st}(\vec u)q_{nm}^{st}(\vec u)}.
\label{add3.4}
\end{equation}
Taking into account that variables in stationary equations are
separated, i.e. $ Q_{nm}^{st}(\vec u)=\delta (u_4)\bar
Q_{nm}^{st}(u_1,u_2,u_3) $ ¨ $ q_{nm}^{st}(\vec u)=\delta (u_4)\bar
q_{nm}^{st}(u_1,u_2,u_3), $ the formula (\ref{add3.44}) for $
C_{nm} $ is obtained.
\eprf


\section{Average transition probabilities in case of
$ \epsilon_1\neq 0 $ and $ \epsilon_2\neq 0 $}
\label{sec-6}

If influence of both random forces is taken into account, i.e. $
\epsilon_1\neq 0 $ and $ \epsilon_2\neq 0, $ evolution of the
system depends on the correlation of the moments $ t_1 $ and $ t_2
$. If $ t<t_>\equiv max(t_1,t_2) $ equations (\ref{102}) and
(\ref{104}) for the trajectories $ \eta (t) $ and $ \xi (t) $ are
different, depending on whether
 $ t_>=t_1 $ or $ t_>=t_2. $ The two possibilities correspond to the two
probability distributions of random variables $ \eta (t_>) $ and $
\xi (t_>) $, which are the initial values for the corresponding
trajectories on the interval $ t>t_>. $ Denote $ z_1=\eta ,\
z_2=\dot \eta ,\ z_3=Re (\dot \xi /\xi ),\ z_4=Im (\dot \xi /\xi ).
$ For the probability density function $ R(\vec z,t) $, which is
defined as a probability for the trajectory $
\vec z(t) $ to be found in the interval $ [\vec z,\ \vec z+d\vec z] $
at the moment $t>t_<\equiv\min(t_1,t_2), $ we can write down the
following expression
\begin{equation}
R(\vec z,t_>)=\left \{
\begin{array}{ll}
P_1(z_1,z_2,t_>\vert z_{01},z_{02},t_<)\delta (z_3-z_{03})
\delta (z_4-z_{04}),& t_>=t_1, \\ \\
P_2(\vec z,t_>\vert \vec z_0,t_<),& t_>=t_2, \\
\end{array} \right.
\label{401}
\end{equation}
where
$$
\vec z_0=(z_{01},z_{02},z_{03},z_{04})=\Big (\eta_0(t_<),
\ \dot \eta_0(t_<),\ Re (\dot \xi_0(t_<)/\xi_0(t_<)),\ Im
(\dot \xi_0(t_<)/\xi_0(t_<)\Big ).
$$
and functions $ P_1 $ and $ P_2 $ are the solutions of (\ref{205})
and (\ref{305}), respectively. It is obvious that the normalization
condition $ \int R(\vec z,t_>) d\vec z = 1 $ holds.

At $ t>t_> $ we have the same set of SDE for the components of the
random vector process $\vec z(t) $ both at $ t_>=t_1 $ and $
t_>=t_2. $ Its inference literally reproduces the derivation of the
set (\ref{304}) and results in
\begin{equation}
\left\{
\begin{array}{ll}
\dot z_1=z_2,&\\
\dot z_2=F_0-\Omega_0^2z_1-\sqrt{2\epsilon_1p_1}z_1f_1+
\sqrt{2\epsilon_2p_2}f_2,& \\
\dot z_4=-2z_3z_4,&\\
\dot z_3=z_4^2-z_3^2-\Omega_0^2(t)-\sqrt{2\epsilon_1p_1}f_1(t),&
\vec z(t_>)=\vec z_>,
\end{array}\right.
\label{402}
\end{equation}
where distribution $ R(\vec z_>,t_>) $ of the components of the
random vector $ \vec z_> $ is given by the formula (\ref{401}). The
representation of the joint probability density
$$
P_3(\vec z,t\vert \vec z_>,t_>)=\Big <\delta (\vec z(t)-
\vec z)\Big >\Bigr \vert_{
\vec z(t_>)=\vec z_>,
}
$$
derived by the standard method from (\ref{402}), is given by the
Fokker-Planck equation. Thus we arrive at the analogue of the
theorem \ref{sec-5}.\ref{theor-4.1}
\begin{theor}
In the case of $ \epsilon_1\neq 0 $ and $ \epsilon_2\neq 0 $ the
set of equations (\ref{102}), (\ref{104}) generates at $ t>t_> $
the set of SDE describing the evolution of four random processes
$$
\vec z(t)\equiv \big (z_1(t),z_2(t),z_3(t),z_4(t)\big )\equiv
\left (\eta (t),\dot \eta (t),Re\left (\frac {\dot \xi (t)}
{\xi (t)}\right ),
Im\left (\frac {\dot \xi (t)}{\xi (t)}\right )\right )
$$
with the joint probability distribution $ P_3(\vec z,t\vert \vec
z_>,t_>),\ t>t_> $ satisfying the Fokker-Planck equation
\begin{equation}
\frac {\partial P_3}{\partial t}=\hat L_3P_3,
\label{403}
\end{equation}
$$
\hat L_3(\vec z)\equiv
-\sum_{i=1}^4
K_i\frac {\partial }{\partial z_i}+(\epsilon_2p_2+
\epsilon_1p_1z_1^2)\frac {\partial^2 }{\partial z_2^2}+
\epsilon_1p_1\frac {\partial^2 }{\partial z_3^2}+
2\epsilon_1p_1z_1\frac {\partial^2 }{\partial z_2
\partial z_3}+4z_3,
$$
$$
K_1=z_2,\quad K_2=F_0-\Omega_0^2z_1,\quad K_3=z_4^2-z_3^2-
\Omega_0^2,\quad K_4=-2z_3z_4.
$$
and the initial condition
$$
P_3\Big \vert_{t=t_>}=\delta (\vec z-\vec z_>),
$$
with the probability distribution of components of the random
vector $\vec z_> $ given by the formula (\ref{401}).
\label{theor-5.1}
\end{theor}
\vspace{0.5cm}

It is not difficult to show that if $\epsilon_1=0 $ the equation
(\ref{403}) transforms into (\ref{205}) by the substitution
$$
P_3(\vec z,t\vert \vec z_>,t_>)=
\delta \left (z_3-Re \frac {\dot \xi_0(t)}{\xi_0(t)}\right )
\delta \left (z_4-Im \frac {\dot \xi_0(t)}{\xi_0(t)}\right )
P(z_1,z_2,t).
$$
When $ \epsilon_2=0 $ the equation (\ref{403}) transforms directly
into (\ref{305}). Thus the relation of the general case with the
particular situations, considered in the previous sections, is
established.

Using again the formulas (\ref{1003})-(\ref{1004}), the
representation (\ref{309}) of $\left <\vert c_{nm}\vert^2\right > $
may be generalized to the case under consideration
\begin{equation}
\left <\vert c_{nm}\vert^2\right >=
\int d\vec z_> R(\vec z_>,t_>)
\int d\vec z H_{nm}(\vec z)Q_{nm}(\vec z,t).
\label{404}
\end{equation}
As compared with (\ref{309}), an additional integration with the
weighting function $ R(\vec z_>,t_>) $ allowing for the dispersion
of initial values of the trajectory, was included into (\ref{404}).
Functions $ H_{nm}(\vec z) $ are defined in (\ref{310}) and
functions $ Q_{nm}(\vec z,t), $ depending on $ z_> $ as on a
parameter, are solutions of the following problem
$$
\frac {\partial Q_{nm}}{\partial t}=\hat L_3Q_{nm}-V_{nm}Q_{nm},
$$
\begin{equation}
Q_{nm}(\vec z,t)
\mathrel{\mathop{\longrightarrow }\limits_{t\rightarrow t_>}}
\delta (\vec z-\vec z_>),
\quad
 Q_{nm}(\vec z,t)
\mathrel{\mathop{\longrightarrow }\limits_{\vert \vert \vec z\vert
\vert \rightarrow \infty }}0,
\label{405}
\end{equation}
where
$$
V_{nm}=p_{nm}z_3,\quad p_{00}=p_{01}=1,\quad p_{10}=p_{11}=3.
$$
Proceeding from (\ref{404})-(\ref{405}) the theorems analogous to
the theorems \ref{sec-5}.\ref{theor-3} and
\ref{sec-5}.\ref{theor-4.3} can be proved. Here we shall combine
them into the one
\begin{theor}
Let $ \Omega_0(t)\equiv const=\Omega_0,\ F_0(t)\equiv const=F_0. $
Let the functions $ p_i(t) $ have the form
\begin{equation}
p_i(t)=
\left \{
\begin{array}{ll}
1&t_1<t<t_e,\\
0&t>t_e.
\end{array}
\right.
\label{add312}
\end{equation}
and let the moment $ t_e $ lie far enough in the future to provide
the legitimacy of the approximate replacement of the solution $
Q_{nm}\left (\vec u,t\right ) $ of the equation (\ref{311}) taken
at this moment by its stationary limit at $t\to
\infty . $ Then we have the following representation for the
probabilities $ W_{nm} $
\begin{equation}
W_{nm}=\Omega_{in}^{p_{nm}}A_{nm}(t_>)
\int d\xi_1d\xi_2d\xi_3
{\bar q_{nm}^{st}(\xi_1,\xi_2,\xi_3)}
\bar H_{nm}\left (\xi_1, \xi_2, \xi_3\right ).
\label{325dop}
\end{equation}
$$
A_{nm}(t_>)\equiv \int d\vec z_> R(\vec z_>,t_>)C_{nm}(\vec z_>)
$$
where $ \bar H_{nm} $ are defined in (\ref{326}), the function $
\bar q_{nm}^{st} $ is an arbitrary (decreasing on infinity)
solution of the shortened stationary equation
\begin{equation}
\left (\hat {\bar L}_3^{st}-p_{nm}\xi_3\right )\bar q_{nm}^{st}=0,
\label{407}
\end{equation}
$$
\hat {\bar L}_3^{st}(\vec \xi )\equiv
-\sum_{i=1}^4\bar K_i\frac \partial {\partial \xi_i}
+
(\epsilon_2+\epsilon_1\xi_1^2)\frac {\partial^2 }{\partial \xi_2^2}+
\epsilon_1\frac {\partial^2 }{\partial \xi_3^2}+
2\epsilon_1\xi_1\frac {\partial^2 }{\partial \xi_2
\partial \xi_3}+
2z_3,
$$
$$
\bar K_1=z_2,\quad \bar K_2=-\Omega_{out}^2z_1,\quad
\bar K_3=-(z_3^2+\Omega_{out}^2),
$$
and the function $ C_{nm}(\vec z_>) $ depending on the parameter $
z_> $ is defined by the formula
\begin{equation}
C_{nm}= {Y_{nm}^{st}(\vec z_>)}
\left \{\int dz_1dz_2dz_3\
Y_{nm}^{st}(\vec z)\Big \vert_{z_4=0}\bar q_{nm}^{st}(z_1,z_2,z_3)\right \}^{-1}
\label{add4.44}
\end{equation}
where $ Y_{nm}^{st}(\vec z) $ is an arbitrary solution of the
conjugate equation
\begin{equation}
\left (\left (\hat {L}_3^{st}\right )^+(\vec z)-p_{nm}z_3\right )
Y_{nm}^{st}(\vec z)=0,
\label{add.adjoint.new}
\end{equation}
$$
\left (\hat {\bar L}_3^{st}\right )^+\left (\vec z\right )\equiv
\sum_{i=1}^4K_i^{st}\frac {\partial }{\partial z_i}+
(\epsilon_2+\epsilon_1z_1^2)\frac {\partial^2 }{\partial z_2^2}+
\epsilon_1\frac {\partial^2 }{\partial z_3^2}+
2\epsilon_1z_1\frac {\partial^2 }{\partial z_2
\partial z_3},
$$
$$
 K_1^{st}=z_2,\quad K_2^{st}=-\Omega_{out}^2z_1,\quad
 K_3^{st}=z_4^2-z_3^2-\Omega_{out}^2),\quad K_4^{st}=-2z_3z_4.
$$
\label{theor.last}
\end{theor}


\section{Another approach to the calculation of average transition probabilities}
\label{sec-7}

There is an alternative approach to calculation of the average
transition probabilities. The main idea of the method will be
demonstrated with the particular case of the problem (\ref{001}) -
(\ref{005}).
\begin{theor}
Let in (\ref{003}) $\Omega_0 = F_0 = 0,\ \epsilon_1 = 0. $ Then the
solution of the problem (\ref{001})-(\ref{003}) may be represented
as
\begin{equation}
\Psi_{stc}(t,x)= e^{i(\xi_1(x-\xi_2)+\xi_3)}\phi (x-\xi_2,t)
\label{902}
\end{equation}
where function $ \phi (y,t) $ satisfies a usual partial
differential equation
$$
i\frac {\partial \phi }{\partial t}=-\frac 1 2
\frac {\partial^2\phi }{\partial^2y}+\frac i 2y^2\phi ,
$$
and random processes $ \xi_1,\ \xi_2,\ \xi_3 $ satisfy the system
of ordinary SDE
\begin{equation}
\left \{
\begin{array}{rcl}
d\xi_1(t)&=&-i\xi_2(t)dt+\sqrt{2\epsilon_2}dW_2(t)
\\
d\xi_2(t)&=&\xi_1(t)dt
\\
d\xi_3(t)&=&\displaystyle \frac 1 2(\xi_1^2(t)-i\xi^2(t))dt+
\sqrt{2\epsilon_2}\xi_2(t)dW_2(t).
\end{array}
\right.
\label{903}
\end{equation}
\end{theor}
\prf
Let us rewrite the equation (\ref{001}) as a SDE \cite{Gard}
\begin{equation}
id\Psi_{stc}= -\frac {1}{2}\frac {\partial^2\Psi_{stc}}{\partial x^2}dt-
\sqrt{2\epsilon_2}\Theta (t-t_2)x\Psi_{stc}dW_2(t),
\label{901}
\end{equation}
which determines an increment of the random process $ \Psi_{stc} $
during the time interval $ dt. $ The quantity $ W_2(t) $ in
(\ref{901}) is the Wiener process $ \left
(W_2(t)=\int_{t_2}^tf_2(t')dt'\right ), $ i.e. the Gaussian random
process which is completely determined by the initial condition $
W_2(t_2)=0 $ and transition probability
$$
P(w_0,t_0\vert w,t) = \frac {1}{\sqrt{2\pi (t-t_0)}}
\exp{\left [-\frac {(w-w_0)^2}{2(t-t_0)}\right ]}.
$$
The equation (\ref{901}) differs from an ordinary SDE in that it
includes derivatives of the process with respect to parameter $ x.
$ The idea of the method is to represent $
\Psi_{stc} $ as a function of several random processes satisfying
the set of ordinary SDE, the latter being chosen so as (\ref{901})
would be turned into an identity after the substitution of $
\Psi_{stc} $. In doing so it is necessary to apply
the Ito calculus as follows. If $
\Psi_{stc}=\phi (\vec
\xi ), \mbox{where}\ \vec \xi =(\xi_1, ... ,\xi_n) $ is a set of
random processes satisfying the set of ordinary SDE
\begin{equation}
d\xi_i(t)=a_i(\vec \xi ,t)dt+b_i(\vec \xi ,t)dW_2(t),\quad
i=1, ... , n,
\label{infi_xi}
\end{equation}
and $ \phi $ is a differentiable function of its arguments, then an
increment of the process $ \Psi_{stc} $ is
$$
d\Psi_{stc}(t)=\left [\sum_{i=1}^n\left (a_i\frac {\partial \phi }
{\partial \xi_i}+\frac {1}{2}b_i^2\frac {\partial^2\phi }{\partial^2
\xi_i^2}\right )+\sum_{i\neq j}b_ib_j\frac {\partial^2\phi}
{\partial \xi_i\partial \xi_j}\right ]dt+
$$
\begin{equation}
+\sum_{i=1}^nb_i
\frac {\partial \phi }{\partial \xi_i}dW_2(t).
\label{infi_Psi}
\end{equation}
where $ d\xi_i $ are infinitely small increments of the processes $
\xi_i $ determined from (\ref{infi_xi}). Using this rule, we can
show after simple but cumbersome manipulations that the equation
(\ref{901}) holds at $ t>t_2 $, if $ \Psi_{stc} $ is represented by
(\ref{902})-(\ref{903}).
\eprf

The initial conditions for $ \xi_i $ may be set to zero:
$
\xi_1(t_2)=\xi_2(t_2)=\xi_3(t_2)=0. $ Then solving the equation
(\ref{901}) at $ t<t_2 $ and denoting the solution as
 $ \Psi_{stc}^<(t,x), $  we obtain the initial condition for
$ \phi $ as: $ \phi (t_2,x)=\Psi_{stc}^<(t_2,x). $ The boundary
conditions for $ \phi $ are imposed by the natural requirement of
normability of the solution.

Having the solution (\ref{902}) it is not difficult to find the
matrix elements of different quantum mechanical operators and the
probabilities of transitions into given stationary states (for
example, to $\phi_n^{out}). $ To be able to average
these quantities over the joint distribution $ P(t,\vec
\xi ), $ we must solve the Fokker-Planck equation corresponding to
the set (\ref{903}).

Unfortunately a substitution reducing the equation (\ref{001}) to
the set of ordinary SDE is unknown in the general case
(\ref{001})-(\ref{005}). Therefore at present the described method
provides noting more than an instructive illustration of the one of
possible approaches to the problem to be solved. Though not leading
to the final results, it reserves the field for the further study.
Undoubtedly it would be very interesting to obtain the solution
with the help of the described procedure, even if only to compare
it to those found in the previous sections.


\section{Thermodynamics within the framework of representation by stochastic
density matrix. Thermodynamical characteristics of oscillator
}

It is well known \cite{Zubar} that the key object of interest in
quantum mechanics is the density matrix.

\begin{defin} The stochastic density matrix is defined by
the expression
\begin{equation}
\label{eq8.1.9} \rho_{stc}\left( x,t;\{ \vec \zeta \} \vert
x^{\prime },t^{\prime };\{ \vec \zeta^{\prime }\} \right) =
\sum_{m=0}^\infty
 w_0^{\left( m\right)
}\rho_{stc}^{\left( m\right) } \left( x,t;\{ \vec \zeta \} \vert
x^{\prime },t^{\prime }; \{ \vec \zeta^{\prime }\} \right) ,
\end{equation}
\begin{equation}
\label{eq8.1.10}
\rho_{stc}^{\left( m\right) }\left( x,t;\{ \vec \zeta \} \vert x^{\prime },
t^{\prime };\{ \vec \zeta^{\prime }\}
\right) =\sqrt{\frac {\pi }{\Omega_{in}}}\Psi_{stc}^{\left( m\right) }
\left( x,t\vert \{ \vec \zeta
\} \right) \overline{\Psi_{stc}^{\left( m\right) }
\left( x^{\prime },t^{\prime}\vert \{ \vec \zeta^{\prime }\} \right) },
\end{equation}
where $w_0^{\left( m\right) }$ has the meaning of the initial
distribution over quantum states with energies $E_m=\left( \frac
12+m\right) \Omega _{in},$ until the moment when the generator of
random excitations is activated.
\end{defin}

\begin{defin} The expected value of the operator
$\hat A\left( x,t\vert \{ \vec \zeta \} \right) $ in quantum state
with the index $m$ is
\begin{equation}
\label{eq8.1.18}
A_m=
\lim\limits_{t\rightarrow +\infty } \left\{ \left. Sp_x\left[
Sp_{\{ \vec \zeta \} }\hat A\rho_{stc}^{\left( m\right) }\right]
\right/ Sp_x\left[ Sp_{\{ \vec \zeta \} }\rho _{stc}^{\left(
m\right) }\right] \right\} .
\end{equation}
The mean value of the operator $\hat A\left( x,t\vert
\{
\vec
\zeta
\} \right) $ over the whole ensemble of states will respectively be
given by
\begin{equation}
\label{eq8.1.18a}
A=
\lim\limits_{t\rightarrow +\infty }\left\{ \left. Sp_x\left[
Sp_{\{ \vec \zeta \} }\hat A\rho _{stc}\right] \right/ Sp_x\left[
Sp_{\{ \vec \zeta \} }\rho _{stc}\right] \right\} .
\end{equation}
\end{defin}
The operation $ Sp_x $ in (\ref{eq8.1.18}) and (\ref{eq8.1.18a}) is
defined by
\begin{equation}
\label{eq8.1.12new_A}
Sp_x\{K(x,x')\}=\sqrt{\frac {\Omega_{in}}{\pi }}\int dx K(x,x)
\end{equation}
for any function $ K(x,x'). $

Using (\ref{eq8.1.10}) and the properties of the functionals $
\Psi_{stc}^{\left( m\right) }\left( x,t\vert \{ \vec \zeta
\}\right) $ we easily obtain the expression for the total
nonstationary distribution function
\begin{equation}
\label{eq8.1.12}
w_0=Sp_xSp_{\{ \vec \zeta \} }\left\{ \rho _{stc}\left( x,t;\{ \vec \zeta \}
\vert x^{\prime },t^{\prime };\{ \vec \zeta^{\prime }\}
\right) \right\}
=
\sum_{m=0}^{\infty }w_0^{\left( m\right) }.
\end{equation}

If the initial weighting functions $w_0^{\left( m\right) }$ are
given by the canonical distribution $ w_0^{\left( m\right) }=\exp
\left (-E_m/{\rm k}T\right ) $, the expression (\ref{eq8.1.12})
takes the form of the Planck distribution (see \cite{Zubar})
\begin{equation}
\label{eq8.1.16}
w_0\left( \beta \right) =\frac{e^{\beta /2}}{e^\beta -1}%
,\qquad \beta =\frac {\Omega_{in}}{{\rm k}T}.
\end{equation}

Substituting the expansion (\ref{razlogh}) of the wave functional
in {\em out}-states into (\ref{eq8.1.9})-(\ref{eq8.1.10}) we have
the following representation
$$
\rho _{stc}\left( x,t;\{ \vec \zeta \} \vert x^{\prime },t^{\prime
};\{ \vec \zeta^{\prime }\} \right) =
$$
\begin{equation}
\label{eq8.1.16new}
=
\sum\limits_{m,k,l=0}^{\infty }w_0^{\left( m\right) }c_{mk}\left(
t\vert \{ \vec \zeta \} \right) \overline{c_{ml}\left( t\vert \{
\vec \zeta \} \right) }\phi_k^{out}(x,t)
\overline{\phi_l^{out}(x',t) }
\end{equation}

\begin{defin} The nonequilibrium partial distribution function is
defined by
$$
w^{\left( m\right) }\left( \epsilon _{1},\epsilon _{2},t\right) =
Sp_{\{ \vec \zeta \} }
\left\{ %
\sum_{k=0}^\infty \left [
w_0^{\left( k\right) }\left|
c_{km}\left( t\vert \{ \vec \zeta \} \right) \right| ^2-
w_0^{\left( m\right) }\left|
c_{mk}\left( t\vert \{ \vec \zeta \} \right) \right| ^2\right ]
\right\}+w_0^{\left( m\right) }=
$$
\begin{equation}
\label{eq8.1.18new}
=\sum_{k=0}^\infty
\left \{ \left [
w_0^{\left( k\right) }
\Delta_{km}\left( t\right) -
w_0^{\left( m\right) }
\Delta_{mk}\left( t\right)\right ]
\right\}+w_0^{\left( m\right) },
\end{equation}
where
$$
\Delta _{km}\left( t\right) =Sp_{\{ \vec \zeta \} }\left|
c_{km}\left( t\vert \{ \vec \zeta \} \right) \right| ^2=\left\langle
\left| c_{mm}\left( t\vert \{ \vec \zeta \} \right) \right|
^2\right\rangle .
$$
\end{defin}
In this case the total distribution function is equal to the sum
\begin{equation}
\label{eq8.1.17new}
w_0=\sum_{m=0}^\infty w^{\left( m\right) }
\left( \epsilon _{1},\epsilon _{2},t\right).
\end{equation}

In case under consideration one can introduce different
definitions for such thermodynamical quantity as an entropy.
Despite formal similarity definitions done may provide or not the
connection of defined quantity with irreversibility of the system
evolution. For example one can define the total and the partial
entropy in the following way.

\begin{defin} The formal total entropy of nonequilibrium state is defined as
\begin{equation}
\label{eq8.1.20new_A}
S\left( \epsilon _1, \epsilon _2,t\right) =-Sp_{\{ \vec \zeta \} }Sp_x\left \{\rho_{stc}
\ln{\rho_{stc}}\right \}.
\end{equation}
\end{defin}

\begin{defin} The formal partial nonequilibrium entropy is defined as
\begin{equation}
\label{eq8.1.21new_A}
S^{(m)}\left( \epsilon _{1}, \epsilon _2,t\right) =-Sp_{\{ \vec \zeta \} }Sp_x\left \{\rho^{(m)}_{stc}
\ln{\rho^{(m)}_{stc}}\right \}.
\end{equation}
\end{defin}
It is not difficult to show that the formal partial entropy does
not depend on time and has no relation to thermodynamical
irreversibility.
\begin{prop}
For any $ m $ the formal partial entropy $ S^{(m)}\left( \epsilon
_{1}, \epsilon _2,t\right) $ is equal to zero.
\end{prop}
\prf Let's consider the $ N $-dimensional square matrix $ \hat A
$ with elements $ \hat A_{ik}=a_ia_k, $ where $ a_i,\ i=1,\dots
,N $ are the elements of $ N $-dimensional vector. It is possible
to find all eigenvalues $ \lambda_i $ and to find out the
structure of eigen-subspaces for matrix $ A. $ Namely, one can
show that $ \lambda_1=a_1^2+a_2^2+\dots +a_N^2, $ $ \
\lambda_2=\lambda_3=\dots =\lambda_N=0. $ At that eigenvector $
\bf e_1 $ coincides with $ \bf a, $ and eigen-subspace
corresponding to zero eigenvalues is orthogonal to $ \bf a. $

Generalizing this result on the case of infinitely dimensional
matrix $ \rho^{(m)}_{stc}, $ one obtains: there is one
eigenvector $ \left ({\pi }/{\Omega_{in}}\right )^{1/4}
\Psi_{stc}^{\left( m\right) }\left( x,t\vert \{ \vec \zeta \}
\right), $ corresponding to nonzero eigenvalue $ \lambda_1=Sp_{\{
\vec \zeta \} }Sp_x\left \{\rho^{(m)}_{stc}\right \}, $ and there
is an infinitely dimensional eigen-subspace, corresponding to
zero eigenvalue, which is orthogonal to this vector.
Supplementing the vector $ \left ({\pi }/{\Omega_{in}}\right
)^{1/4} \Psi_{stc}^{\left( m\right) }\left( x,t\vert \{ \vec
\zeta \} \right) $ with any orthonormal set of vectors lying in
the subspace mentioned above, one obtains the basis of the whole
space which brings matrix $ \rho^{(m)}_{stc} $ to diagonal form.
Understanding the uncertainty $ 0\ln{0} $ as a limit
$$
0\ln{0}=\lim_{s\to 0}{s\ln{s}}=0,
$$
one obtains for formal partial entropy:
$$
S_f^{(m)}\left( \epsilon _{1}, \epsilon _2,t\right)= Sp_{\{ \vec
\zeta \} }Sp_x\left \{\rho^{(m)}_{stc}\right \}\cdot Sp_{\{ \vec
\zeta \} }Sp_x\left \{\ln{\rho^{(m)}_{stc}}\right \}=0,
$$
which makes the proof complete. \eprf

If one wishes to have the quantity describing irreversible
behavior of the system, it is necessary to change definition of
entropy.
\begin{defin} Total and partial entropies of nonequilibrium
state are defined as
\begin{equation}
\label{eq8.1.20new_A1} S\left( \epsilon _1, \epsilon _2,t\right)
=-Sp_x\left \{\rho_{av} \ln{\rho_{av}}\right \}.
\end{equation}
and
\begin{equation}
\label{eq8.1.21new_A1} S^{(m)}\left( \epsilon _{1}, \epsilon
_2,t\right) = -Sp_x\left
\{\rho^{(m)}_{av}\ln{\rho^{(m)}_{av}}\right \}
\end{equation}
correspondingly, where
$$
\rho_{av}=Sp_{\{ \vec \zeta \} }\left \{\rho_{stc}\right \},\quad
\rho^{(m)}_{av}=Sp_{\{ \vec \zeta \} }\left
\{\rho^{(m)}_{stc}\right \}.
$$
\end{defin}
Unfortunately we have no at the moment simple enough analitical
representation for the quantities defined in such a way.

To illustrate the definitions given above we calculate the
average energy of oscillator in the ground, vacuum, state (i.e.
at $m=0$) assuming that both regular and stochastic parts of the
external force are absent. In this case the density matrix has
the form
$$
\rho _{stc}^{\left( 0\right) }\left( x,t;\{ \vec \zeta \} \mid
x^{\prime },t^{\prime };\{ \vec \zeta^{\prime }\} \right)
=\exp \left\{ -\frac{\Omega _{in}}2(x^2+x^{\prime
2})-\right.
$$
\begin{equation}
\label{eq8.2.1} \left. -\frac 12
\int\limits_{t_{1}}^t \Phi \left( \tau \right) d\tau -\frac 12
\int\limits_{t_{1}}^{t'}\Phi^{*}\left( \tau \right) d\tau -\frac
i2\left[ \Phi \left( t\right) x^2-\Phi^{*}\left( t^{\prime
}\right) x^{\prime 2}\right] \right\} .
\end{equation}


\begin{prop}
Let $ \Omega (t)\equiv \Omega_{in}, $ $ F_0(t)\equiv 0, $ $
\epsilon_2=0, $ $ p_1(t)\equiv 1. $
Then the average energy
$$
E_{osc}^{(0)}(\lambda )=Sp_xSp_{\{ \vec \zeta \} }
\left (\hat H\tilde \rho_{stc}^{(0)}\right )
$$
is represented by
\begin{equation}
\label{eq8.2.3}
E^{(0)}_{osc}\left( \lambda \right) =\frac 12\Omega
_{in}\left\{ 1-\frac 1{\sqrt{\lambda }}
\int\limits_{-\infty }^{+\infty }
d\bar u_3 K_1\left( \lambda
,\bar u_3\right)
+\frac i{\sqrt{\lambda }}
\int\limits_{-\infty }^{+\infty }
d\bar u_3 K_2\left( \lambda
,\bar u_3\right) \right\} ,
\end{equation}
with the designations
$$
K_1\left( \lambda ,\bar u_3\right) =
C_0\bar u_3 \bar q^{st}\left( \lambda ,\bar
u_3\right) \times
$$
$$
\times
\left\{ -\sqrt{\frac{A_{00}-1}{2A_{00}^2}}+\frac{\bar u_3}
{\sqrt{\lambda }A_{00}^2}\left[ \sqrt{\frac{A_{00}+1}{2A_{00}^2}}+%
\frac{\bar u_3}{\sqrt{\lambda }}\sqrt{\frac{A_{00}-1}{2A_{00}^2}}%
\right]
\right\} ,
$$
$$
K_2\left( \lambda ,\bar u_3 \right) =
C_0\bar u_3 \bar q^{st}\left( \lambda ,\bar
u_3 \right) \times
$$
$$
\left\{ -\sqrt{\frac{A_{00}-1}{%
2A_{00}^2}}+\frac{\bar u_3}{\sqrt{\lambda }A_{00}^2}\left[ -\sqrt{%
\frac{A_{00}-1}{2A_{00}^2}}+\frac{\bar u_3 }{\sqrt{\lambda }}\sqrt{%
\frac{A_{00}-1}{2A_{00}^2}}\right] \right\} ,
$$
\begin{equation}
\label{eq8.2.4}
A_{00}\left( \lambda ,\bar u_3\right) =\sqrt{1+%
\frac{\bar u_3^2}{\lambda }}.
\end{equation}
The function $ q^{st}(u_3) $ is an arbitrary solution of the
equation
\begin{equation}
\epsilon_1\frac {d^2q^{st}}{du_3^2}+(u_3^2+\Omega_{in}^2)
\frac {dq^{st}}{du_3}+u_3q^{st}=0,
\label{dop2}
\end{equation}
decreasing as $ \vert u_3\vert \to \infty ,$ and the constant $ C_0
$ given by
\begin{equation}
C_0=\frac {Y^{st}(u_{30},u_{40})}{\int du_3\ Y^{st}(u_3,0)q^{st}(u_3)}.
\label{dop3}
\end{equation}
Here the function $ Y^{st}(u_3,u_4) $ is an arbitrary solution of
the equation
\begin{equation}
\epsilon_1\frac {d^2Y^{st}}{du_3^2}-(u_3^2-u_4^2+\Omega_{in}^2)
\frac {dY^{st}}{du_3}-2u_3u_4\frac {dY^{st}}{du_4}-u_3Y^{st}=0,
\label{dop4}
\end{equation}
decreasing as $ u_3^2+u_4^2\to \infty .$
\end{prop}
\prf
In fact the proof copies the manipulations performed repeatedly in
this paper and thus may be omitted.
\eprf

The second term inside the figure brackets in (\ref{eq8.2.3}) is a
level shift which is well known from quantum electrodynamics as the
Lamb shift, the third term determines the magnitude of the ground
state energy broadening. Note that the lifetime at this level is
proportional to the inverse of the broadening
\begin{equation}
\label{eq8.2.5}
\Delta t^{(0)}\sim 2\frac{\sqrt{\lambda }}{\Omega _{in}}%
\left\{
\int\limits_{-\infty }^{+\infty }
d\bar u_3 K_2\left( \lambda
,\bar u_3 \right) \right\} ^{-1}.
\end{equation}

The average energy of a randomly wandering (QHO) for any quantum
level is calculated quite similarly.


\section*{Conclusion}


There are three different reasons which may cause a chaos in the
basic quantum mechanical object, i.e. the wave function. The first
reason refers to measurements performed over a quantum system
\cite{gp,sinai}. The second reason consists in the more fundamental
openness of any quantum system resulting from the fact that all the
beings are immersed into a physical vacuum \cite{glimm}. In the
third place, as it follows from the recent papers
\cite{avt1,avt2,avt3}, a chaos may also appear in the wave function
even in a closed dynamical system. As it is shown in \cite{sinai},
there is a close connection between a classical nonintegrability
and a chaos in the corresponding quantum system. Many of the
fundamentally important questions of the quantum physics such as
the Lamb shift of energy levels, spontaneous transitions between
the atom levels, quantum Zeno effect \cite{itano}, processes of
chaos and self-organization in quantum systems, especially those
where the phenomena of phase transitions type may occur, can be
described qualitatively and quantitatively in a rigorous way only
within the nonperturbative approaches. The Lindblad representation
\cite{gorini,lindblad} for the density matrix of the system
"quantum object + thermostat" describes {\em a priori} the most
general situation which may appear in the nonrelativistic quantum
mechanics. Nevertheless, we need to consider a reduced density
matrix on a semi-group \cite{gp}, when investigating a quantum
subsystem. This is quite an ambiguous procedure and moreover its
technical realization is possible only in the framework of a
particular perturbative scheme.

A crucially new approach to constructing the quantum mechanics of
the closed nonrelativistic system "quantum object + thermostat" has
been developed recently by the authors of \cite{avt1,avt2} from the
principle of "local correctness of Schr\"{o}dinger representation".
To put it differently, it has been assumed that the evolution of
the quantum system is such that it may be described by the
Schr\"{o}dinger equation on any small time interval, while the
motion as a whole is described by a SDE for the wave function. In
this case, however, there emerges not a simple problem to find a
measure for calculating the average values of the physical system
parameters. Nevertheless, there exists a certain class of models
for which all the derivations can be made not applying the
perturbation theory \cite{avt2}.

In the present paper we explore further the possibility of building
the nonrelativistic quantum mechanics of closed system "quantum
object + thermostat" within the framework of the model of
one-dimensional randomly wandering QHO (with a random frequency and
subjected to a random external force). Mathematically the problem
is formulated in terms of SDE for a complex-valued probability
process defined on the extended space $ R^1\otimes R_{\{ \xi \} }.
$ The initial SDE is reduced to the Schr\"{o}dinger equation for an
autonomous oscillator defined on a random space-time continuum,
with the use of a nonlinear transformation and one-dimensional
etalon nonlinear equation of the Langevin type defined on the
functional space $ R_{\{ \xi\} }. $ It is possible to find for any
fixed $ \{ \xi\} $ an orthonormal basis of complex-valued random
functionals in the space $ L_2(R^1) $ of square-integrable
functions. With the assumption that the random force generator is
described by a white noise correlator, the Fokker-Planck equation
for a conditional probability is found. From the solutions of this
equation on an infinitely small time interval a measure of the
functional space $ R_{\{ \xi \} }$ can be constructed. Then by
averaging an instantaneous value of the transition probability over
the space $ R_{\{\xi \} } $, the mean value of the transition
probability is represented by a functional integral. Using the
generalized Feynman-Kac theorem, it is possible to reduce the
functional integral in the most general case, where both frequency
and force are random, to a multiple integral of the fundamental
solution of some parabolic partial differential equation. The
qualitative analysis of the parabolic equation shows that it may
have discontinuous solutions \cite{josef}. This is equivalent to
the existence of phenomena like the phase transitions in the
microscopic transition probabilities. In the context of the
developed approach the representation of the stochastic density
matrix is introduced, which allows to build a closed scheme for
both nonequilibrium and equilibrium thermodynamics. The analytic
formulas for the ground energy level broadening and shift are
obtained, as well as for the entropy of the ground quantum state.

The further development of the considered formalism in application
to exactly solvable many-dimensional models may essentially extend
our understanding of the quantum world and lead us to the new
nontrivial discoveries.



\section*{Appendix}

\setcounter{equation}{0}
\renewcommand{\theequation}{A.\arabic{equation}}

\begin{theor_append}
Let the set of random processes $ (\xi_1, \xi_2,..,\xi_n)\equiv
\vec \xi $  satisfy the set of SDE
$$
\dot \xi_i(t)=a_i\left (\vec \xi ,t\right )+\sum_{j=1}^n
b_{ij}\left (\vec \xi ,t\right )f_j(t),\quad
i=1, ... , n,
$$
$$
\left <f_i(t)f_j(t')\right >=\delta_{ij}\delta (t-t'),
$$
so that the Fokker-Planck equation for the conditional transition
probability density
\begin{equation}
P^{(2)}\left (\vec \xi_2,t_2\vert \vec \xi_1,t_1\right )=\left <
\delta \left (\vec \xi (t_2)-\vec \xi_2\right )\right >\left
\vert_{\vec \xi (t_1)=\vec \xi_1}\right. \quad t_2>t_1
\label{1000}
\end{equation}
is given by
\begin{equation}
\frac {\partial P^{(2)}}{\partial t}=-\sum_{i=1}^n
\frac {\partial }{\partial \xi_i}\left (a_iP^{(2)}\right )+
\sum_{i,j,l,n}\frac {\partial }{\partial \xi_i}
\left (b_{li}\frac {\partial }{\partial \xi_n}\left
(b_{nj}P^{(2)}\right )\right )
\equiv \hat L^{(n)}(\vec \xi )P^{(2)}.
\label{1001}
\end{equation}
The processes $ \xi_i $ are assumed to be markovian and satisfy the
condition $ \vec \xi (t_0)=\vec \xi_0. $ At the same time the
function (\ref{1000}) gives their exhaustive description:
\begin{equation}
P^{(n)}\left (\vec \xi_n,t_n;... \vec \xi_1,t_1; \vec \xi_0,t_0\right )=
P^{(2)}\left (\vec \xi_n,t_n\vert \vec \xi_{n-1},t_{n-1}\right )...
P^{(2)}\left (\vec \xi_1,t_1\vert \vec \xi_0,t_0\right )
\label{1002}
\end{equation}
where $ P^{(n)} $ is a density of the probability that the
trajectory
$
\vec \xi (t) $ would pass through the sequence of intervals $ \left
[\vec\xi_1, \vec \xi_1+d\vec \xi_1\right ],\ldots , \left [\vec
\xi_n, \vec \xi_n+d\vec \xi_n\right ] $ at the subsequent moments
of time $ t_1<\ldots <t_n $, respectively.

Under these assumptions we can obtain the following representation
for an averaging procedure
\begin{equation}
\left <\exp\left \{-\int_{t_0}^tV_1\left (\vec \xi (\tau ),\vec \xi (t)\right )
d\tau
-V_2\left (\vec \xi (t)\right )\right \} \right >=
\int d\vec \xi e^{-V_2\left (\vec \xi ,t\right )}
Q\left (\vec \xi ,\vec \xi ,t\right ),
\label{1003}
\end{equation}
where $ d\vec \xi =d\xi_1...d\xi_n, $ and the function $ Q\left
(\vec
\xi,\vec \xi ',t\right ) $ is a solution of the problem
\begin{eqnarray}
\frac {\partial Q}{\partial t}=\left [\hat L^{(n)}(\vec \xi )-
V_1(\vec \xi ,\vec \xi ')\right ]Q,\qquad \qquad
\label{1004}
\\
Q\left (\vec \xi ,\vec \xi ',t\right )
\mathrel{\mathop{\longrightarrow }\limits_{t\rightarrow t_0}}
\delta \left (\vec \xi -\vec \xi_0\right ),
\quad
 Q\left (\vec \xi ,\vec \xi ',t\right )
\mathrel{\mathop{\longrightarrow }\limits_{\vert \vert \vec \xi \vert
\vert \rightarrow \infty }}0,
\nonumber
\end{eqnarray}
where $ \vert \vert \cdot \vert \vert $ is a norm in $ R^n. $
\end{theor_append}
\prf
The proof is performed formally under the assumption that all the
manipulations are legal. Denote the left side of the equality
(\ref{1003}) by $ I $ and expand the averaging quantity into the
Tailor series:
\begin{equation}
I=\sum_{n=0}^\infty \frac {(-1)^n}{n!}\mu_n(t),
\label{1005}
\end{equation}
where
$$
\mu_n(t)=\left <\left \{
\stackrel{t}{\mathrel{\mathop{\int }\limits_{t_0}}}
V_1(\tau )d\tau+V_2(t)\right \}^n
\right >=
$$
\begin{equation}
=\sum_{m=0}^n\frac {n!}{m!(n-m)!}
\left <V_2^{n-m}(t)\left \{
\stackrel{t}{\mathrel{\mathop{\int }\limits_{t_0}}}
V_1(\tau )d\tau \right \}^m
\right >=
\label{1006}
\end{equation}
$$
=\sum_{m=0}^n\frac {n!}{(n-m)!}
\left <V_2^{n-m}(t)
\stackrel{t}{\mathrel{\mathop{\int }\limits_{t_0}}}
d\tau_m
\stackrel{\tau_m}{\mathrel{\mathop{\int }\limits_{t_0}}}
d\tau_{m-1}\ldots
\stackrel{\tau_2}{\mathrel{\mathop{\int }\limits_{t_0}}}
d\tau_1
V_1(\tau_m)\ldots V_1(\tau_1) \right >\quad \quad \quad
$$
For brevity sake in (\ref{1006}) there was introduced the
designation
$V_1(\tau )\equiv V_1
\left (\vec \xi (\tau ),\vec \xi (t)\right ), $ $\ V_2(t)\equiv
V_2\left (\vec \xi (t)\right ). $ Using the Fubini theorem, we can
represent the averaging procedure in (\ref{107}) as an integration
with the weight $ P^{(n)} $ from (\ref{1002})
$$
\left <V_2^{n-m}(t)
\stackrel{t}{\mathrel{\mathop{\int }\limits_{t_0}}}d\tau_m
\stackrel{\tau_m}{\mathrel{\mathop{\int }\limits_{t_0}}}d\tau_{m-1}\ldots
\stackrel{\tau_2}{\mathrel{\mathop{\int }\limits_{t_0}}}d\tau_1
V_1(\tau_m)\ldots V_1(\tau_1) \right >=
$$
$$
=\int d\vec \xi
\int d\vec \xi_m\ldots \int d\vec \xi_1
\stackrel{t}{\mathrel{\mathop{\int }\limits_{t_0}}}d\tau_m\ldots
\stackrel{\tau_2}{\mathrel{\mathop{\int }\limits_{t_0}}}d\tau_1
P^{(2)}\left (\vec \xi ,t\Big \vert \vec \xi_m,t_m\right )
P^{(2)}\left (\vec \xi_m,t_m\Big \vert \vec \xi_{m-1},t_{m-1}\right )\ldots
$$
$$
\phantom{\stackrel{\tau_2}{\mathrel{\mathop{\int }\limits_{t_0}}}d\tau_1}
\ldots P^{(2)}\left (\vec \xi_2,t_2\Big \vert \vec \xi_1,t_1\right )
V_2^{n-m}\left (\vec \xi \right )V_1\left (\vec \xi_m,\vec \xi \right )\ldots
V_1\left (\vec \xi_1,\vec \xi \right ).
$$
Changing, where it is necessary, the order of integration we can
obtain the following representation for the $ n $-th moment $
\mu_n(t):
$
\begin{equation}
\mu_n(t)=\sum_{m=0}^n\frac {n!}{(n-m)!}
\int d\vec \xi V_2^{n-m}\left (\vec \xi \right )
Q_m\left (\vec \xi ,\vec \xi ',t\right ),
\label{1007}
\end{equation}
where the countable set of functions $ Q_m\left (\vec \xi ,\vec \xi
',t\right ) $ is determined from the recurrence relations
\begin{equation}
Q_m\left (\vec \xi ,\vec \xi ',t\right )=
\stackrel{t}{\mathrel{\mathop{\int }\limits_{t_0}}}d\tau
\int d\vec \eta \ V_2^{n-m}\left (\vec \xi \right )
P^{(2)}\left (\vec \xi ',t\Big \vert \vec \eta ,\tau \right )
V_1\left (\vec \eta ,\vec \xi '\right )
Q_{m-1}\left (\vec \eta ,\vec \xi ',\tau \right ),
\label{1008}
\end{equation}
$$
m=0,1,2,...,
$$
where
\begin{equation}
Q_0\left (\vec \xi ,\vec \xi ',t\right )=
P^{(2)}\left (\vec \xi ,t\Big \vert \vec \xi_0 ,t_0\right )
\label{1009}
\end{equation}
i.e. in fact the function $ Q_0 $ is independent of $ \vec \xi '. $
Upon the substitution of (\ref{1007}) into (\ref{1005}) we insert
the summation procedure under the integration sign and then,
changing the order of double summation, get the expression
\begin{equation}
I=\int d\vec \xi \ e^{-V_2\left (\vec \xi ,t\right )}
Q\left (\vec \xi ,\vec \xi ,t\right ),
\label{1010}
\end{equation}
where
\begin{equation}
Q\left (\vec \xi ,\vec \xi ',t\right )=\sum_{n=0}^\infty (-1)^n
Q_n\left (\vec \xi ,\vec \xi ',t\right ).
\label{1011}
\end{equation}
The representation (\ref{1003}) is thus obtained.

It remains to prove that the function $ Q $ from (\ref{1010}) is a
solution of the problem (\ref{1004}). Using (\ref{1011}) and
(\ref{1008}) we can easily show that $ Q $ satisfies the integral
equation
\begin{equation}
Q\left (\vec \xi ,\vec \xi ',t\right )+
\stackrel{t}{\mathrel{\mathop{\int }\limits_{t_0}}}d\tau
\int d\vec \eta \
P^{(2)}\left (\vec \xi ,t\Big \vert \vec \eta ,\tau \right)
V_1\left (\vec \eta ,\vec \xi '\right )
Q\left (\vec \eta ,\vec \xi ',\tau \right )=
Q_0\left (\vec \xi ,t\right )
\label{1012}
\end{equation}
Taking into account that $\ Q_0\ $ satisfies (\ref{1001}) and the
initial condition $ \ Q_0\left (\vec \xi ,t_0\right )=\delta \left
(\vec \xi -
\vec \xi_0\right ) $ and is an integrable function, it can be
deduced
from (\ref{1012}) that $ Q $ coincides with the solution of the
problem (\ref{1004}). The representation (\ref{1003}), (\ref{1004})
is thus obtained.
\eprf

\end{document}